\documentclass[10pt,journal]{IEEEtran}

\pdfoutput=1

\usepackage{amsmath,epsfig}
\usepackage{amssymb}
\usepackage{amsfonts}
\usepackage{array,dcolumn}
\usepackage{psfrag}
\usepackage{graphicx}
\usepackage{psfrag}
\usepackage{amsmath}
\usepackage{amsfonts}
\usepackage{amssymb}
\usepackage{accents}
\usepackage{cite}
\usepackage{pstricks,exscale}
\usepackage{float}
\usepackage{verbatim}  
\usepackage{epstopdf}
\usepackage{graphicx}
\usepackage[caption=false]{subfig}

\newtheorem{lemma}{Lemma}

\usepackage{footnote}

\usepackage{fancyhdr}

\graphicspath{{figures/}}

\begin{document}

\title{On Energy-efficiency in Wireless Networks: A Game-theoretic Approach to Cooperation Inspired by Evolutionary Biology}

\author{Zoran~Utkovski$^{* \dagger}$,~\IEEEmembership{Member,~IEEE,}
		Andrej Gajduk$^*$,~Lasko Basnarkov$^{* \#}$,~Darko Bo\v{s}nakovski$^\ddagger$         
        and~Ljupco Kocarev$^{* \# \S}$,~\IEEEmembership{Fellow,~IEEE}
\thanks{$^*$ Macedonian Academy of Sciences and Arts.}
\thanks{$^\#$ Ss. Cyril and Methodius University, Faculty of Computer Science and Engineering, P.O. Box 393, 1000 Skopje.}
\thanks{$^\dagger$ Goce Delcev University, Faculty of Computer Science, Stip.}
\thanks{$\ddagger$ Goce Delcev University, Faculty of Medical Sciences, Stip.}
\thanks{$^\S$ BioCircuits Institute, University of California at San Diego.}
\thanks{Part of the material in this paper will be presented at PHYSCOMNET 2014, adjunct workshop to IEEE WiOpt 2014.} 
\thanks{This work was supported in part by the "Deutsche Forschungsgemeinschaft" (DFG), via the project Li 659/13-1.}
}

\markboth{{\tiny This work has been submitted to the IEEE for possible publication. Copyright may be transferred without notice, after which this version may no longer be accessible}}%
{}

\maketitle

\begin{abstract}
We develop a game-theoretic framework to investigate the effect of cooperation on the energy efficiency in wireless networks. We address two examples of network architectures, resembling ad-hoc network and network with central infrastructure node. Most present approaches address the issue of energy efficiency in communication networks by using complex algorithms to enforce cooperation in the network, followed by extensive signal processing at the network nodes. Instead, we address cooperative communication scenarios which are governed by simple, evolutionary-like, local rules, and do not require strategic complexity of the network
nodes. The approach is motivated by recent results in evolutionary biology which suggest that
cooperation can emerge in Nature by evolution, i.~e. can be favoured by natural
selection, if certain mechanism is at work. As result, we are able to show by experiments that cooperative behavior can indeed emerge and persist in wireless networks, even if the behavior of the individual nodes is driven by selfish decision making. The results from this work indicate that uncomplicated local rules, followed by simple fitness evaluation, can promote cooperation and generate network behavior which
yields global energy efficiency in certain wireless networks.
\end{abstract}

\section{Introduction}
The dramatic increase in the number of users and growth in data
traffic in the past years pose a significant challenge on all
aspects of current communication networks including capacity,
signal processing, complexity, and energy consumption. Current
communication networks will not be able to answer these demands
in near future, which calls for a paradigm shift in the design of network architectures, communication strategies and signal processing tools. The provisioning of energy efficient protocols and communication
schemes is one of the main challenges in the design of present and
future communication networks. The concept of energy efficiency is
particularly relevant to emerging heterogeneous networks which,
besides the "classical" communication nodes, include various other
devices with low-power capabilities, such as sensors and other
nodes producing machine-type traffic~\cite{bandyopadhyay2005spatio}. 

Several recent works show that user cooperation is
of fundamental value for increasing both the network throughput
and the energy efficiency. The study of the fundamental limits of
wireless networks suggests that cooperation among the units can
both decrease energy consumption and reduce interference. In this
context, techniques such as cooperative diversity
\cite{sendonaris2003user1, sendonaris2003user2} and interference
alignment \cite{cadambe2008} have been proposed. Energy efficiency of wireless networks has also been studied in
\cite{zhao2005energy,el2006optimal} under the assumption that all
nodes are interested in minimizing the overall energy consumption
of the network.  The globally
optimal solution, as characterized by the authors, is achieved when the network nodes establish cooperation by relaying packets for other
users. 


Most of the present approaches which deal with the aspects of
cooperative communications, assume that the network nodes act in a
pre-defined way, i.~e. their behavior is determined by (usually)
centralized network rules \cite{zhong2003sprite,buchegger2003wiopt,liu2003reputation,
anantvalee2007reputation,mundinger2008analysis}. This approach leaves no freedom to the
individual nodes to decide about their involvement in the
cooperative act. Since cooperation is associated with a cost
(usually energy) and requires certain signal processing
capabilities (computational complexity), this approach may lead to
a "cooperation burden" which can be unreasonably high for some
network nodes.

While this "centralized" approach is reasonable in networks with
central infrastructure, it is also (somewhat surprisingly) widely
adopted in decentralized networks such as ad-hoc networks. One
important group of these efforts focuses on designing high-level
protocols that prevent users from misbehaving and/or provide
incentives for cooperation. To prevent misbehavior, several
protocols based on reputation propagation have been proposed in
the literature, e.g.,
\cite{zhong2003sprite,buchegger2003wiopt,liu2003reputation,
anantvalee2007reputation,mundinger2008analysis,marbach2005cooperation}. One general
observation is that the proper functioning of these networks is
generally maintained either by enforcing cooperation, or by
keeping track of the cooperative behavior which demands intensive
computation. Other works have used ideas from micro-economy to
construct protocols that reward cooperation\cite{buttyan2001nuglets}. 
 Overall, these protocols are based
on ideas rooted in game theory, but, in most cases are hard to
analyze, due to the complicated underlying network models.

Another group of recent works analyzes energy efficiency from a game theoretic perspective, e.g.
\cite{srinivasan2005analytical,felegyhazi2006nash}, where at each time slot, a certain number of nodes are randomly chosen and assigned to
serve as relay nodes on the source-destination route. 
In
\cite{felegyhazi2006nash}, the authors study the Nash 
equilibrium of packet forwarding in a static network and derive the equilibrium
conditions for both cooperative and non-cooperative strategies. The works in this thrust
utilize the repeated game formulation, where cooperation among
users is sustained by punishment for deviating from the
cooperation point. In \cite{lai2008cooperation}, the authors
consider wireless networks consisting of both selfish and altruistic nodes. They establish the critical role of
the altruistic nodes in encouraging cooperation and elaborate on the sub-optimality of relaying
strategies which ignore the game theoretic aspect.

While we also adopt a game-theoretic framework to study energy efficiency, our work differs in several important aspects. First, we do not
focus on enforcing cooperation, for example by
keeping track of the cooperative behaviour of the users and/or
using punishment and reward policies. In addition, we confine our
strategies to the physical layer and avoid introducing elements,
like virtual currency, which may add significant complexity to the
higher layers. Finally, we do not assume presence of altruistic
nodes in the network, but rather assume that all nodes are selfish
in the sense that they try to minimize their \textit{individual}
energy consumption.

The main essence of the work is that we study cooperative
communication scenarios based on \textit{simple local rules} which
mimic evolutionary principles. The approach is motivated by recent
results in evolutionary biology which suggest that, if certain
mechanism is at work, cooperation can be favored by natural
selection, i.~e. even selfish actions 
of the individual nodes can lead to emergence of cooperative
behavior in the network. From system point of view, one of the
main features of our approach is that we shift from the well
accepted paradigm of "engineered" system design, where the system components have known functions and  
designers maintain separation of concerns. Instead, we look at networks such as
biological, social and economic, which evolve over time as a
result of the interactions between the system entities and the
environment. The motivation behind is the remarkable energy efficiency, information storage and
processing capabilities of living organisms, as compared to present communication systems. Based on these observations, we address the mechanisms which lead
to the emergence of cooperation in wireless networks and discuss
the analogies with evolutionary biology. The results indicate that uncomplicated local
rules, followed by simple fitness evaluation, can generate network
behavior which yields global energy efficiency.

The rest of the paper is organized as follows. In
Section~\ref{sec:energy_efficiency_and_cooperation} we present the
relation between energy efficiency and cooperation and discuss
analogies with biological systems. In
Section~\ref{sec:network_model} we describe the network model and
the studied architectures. In Section
\ref{sec:game_theory} we present a game-theoretic framework for
the study of energy-efficiency in wireless networks. The results are presented in
Section~\ref{sec:experiment_setup}. Section~\ref{sec:discussion} interprets the obtained results and discusses possible implications on systems other than wireless networks. Section~\ref{sec:conclusion}
concludes the paper and presents directions for future work.

\section{Cooperation in Communication Networks: Relations with Biological Systems}
\label{sec:energy_efficiency_and_cooperation}
\subsection{Biological systems} 
Cooperation has played a fundamental role in
many of the major transitions in biological evolution and is
essential to the functioning of a large number of biological
systems \cite{axelrod1981evolution, nowak2006, nowak2006five,
cremer2012}. Observations show that cooperative interactions are
required for many levels of biological organization ranging from
single cells to groups of animals. Human society, as well, is
based to a large extent on mechanisms that promote cooperation.

\begin{figure}[!h]
\begin{center}
\includegraphics[scale=0.17]{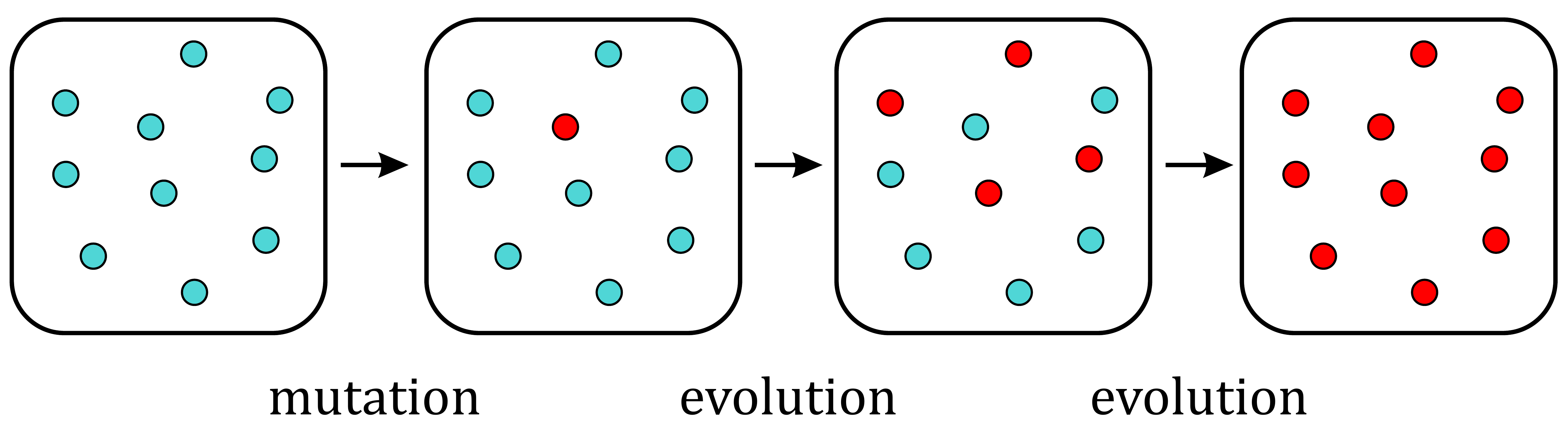}
\caption{Natural selection favours defectors (red) over cooperators (blue), if no other mechanism is at work; After a random defector is introduced in a network of cooperators, cooperation vanishes over time. Source: \cite{nowak2006five}.}
\label{fig:cooperation_extinction}
\end{center}
\end{figure}

Recent results in evolutionary biology suggest that cooperation
can emerge and persist in evolving systems, i.~e. cooperation
can be favored by natural selection, if certain mechanism is
at work \cite {nowak2006, nowak2006five}. These results may be
counter-intuitive since it is well known that in unstructured
populations, natural selection favors defectors over cooperators, as shown in Fig.~\ref{fig:cooperation_extinction}.

However, while studying evolutionary games in structured
populations and on graphs, in \cite{nowak2006} the authors observe
that in structured populations cooperation may emerge given that a
certain mechanism is at work. The approach of capturing this
effects is evolutionary graph theory, which allows the study of
how spatial structure affects evolutionary dynamics. According to
this model, the individuals of a population 
are assumed to be plain cooperators and defectors without
any strategic complexity. The authors show that natural selection
can indeed favor cooperation, if the benefit of the altruistic
act, $b$, divided by the cost, $c$, exceeds the average number of
neighbors, $k$, $b/c>k$. This simple rule is a good approximation
for different graphs, including cycles, spatial lattices, random
regular graphs, random graphs and scale-free networks. In this
setting, the experiments show that cooperators can prevail by
forming network clusters, where they help each other. The
resulting \textit{network reciprocity} is a generalization of
\textit{spatial reciprocity} \cite{nowak2006five}. The intuition
behind is that in this case cooperation can evolve as a
consequence of "social viscosity" even in the absence of
reputation effects or strategic complexity.
It is worth mentioning that, besides network reciprocity, there
are several other candidate mechanisms in biology which are able
to explain the emergence and stability of cooperation in certain
biological systems, such as \textit{kin selection}, \textit{direct
reciprocity}, \textit{indirect reciprocity}, and \textit{group
selection} \cite{axelrod1981evolution,nowak2006five}. As we will
elaborate later on, among these candidate mechanisms we identify
network reciprocity as most relevant to our networks of
interest-wireless communication networks.



\subsection{Wireless Networks: Main Features of the Approach }

Wireless communication systems have two fundamental properties.
The first one is that the receive power decays according to a
power law function of the distance between the users, which puts
stress on the power consumption; the second one is the broadcast
nature of the wireless communication, which leads to interference
between the users. With the increase in the number of subscribers
and growth in data traffic in wireless networks, these two
features gain on importance and have a strong adverse effect on
the network performance in terms of throughput and energy
consumption.

The study of the fundamental limits of wireless networks suggests
that cooperation among the units could potentially overcome these
effects. However, one of the main drawbacks in the performance
analysis of general wireless networks is that it is often based on
simplifying assumptions. As an example, when deriving the
performance limits of different cooperative schemes, the cost of
establishing cooperation in the wireless network is not properly
taken into account \cite{lozano2012}. As result, in some
scenarios, the benefits of cooperation might be overshadowed by
the cost of establishing cooperation in the first place.
 As cooperation comes at a cost
for the network users, in a network which lacks centralized
control, for some users it might be
beneficial to defect, instead of to cooperate. 


Many of the most fundamental instances of cooperation in
biological systems involve simple entities which lack strategic
complexity, which prevents them to adopt strategies that take into
account the history of their interactions with other entities.
Yet, remarkably, cooperation is present in these systems, as
supported by evidence
\cite{soares2008cleaning,mehdiabadi2006social,faaborg1995confirmation}. Similarly, we will be interested in designing rules which are
simple enough to be implemented even by network nodes with
limited processing capabilities, yet powerful enough to promote
cooperation and yield energy efficiency, which is in contrast to the approaches which rely on complex algorithms
and reputation tables in order to enforce cooperation in the
network
\cite{zhong2003sprite,buchegger2002performance,michiardi2002core}.

Essentially, we are interested in cooperation which emerges as 
result of the system evolution.  
The main question we try to answer is the following: Can
cooperation arise in communication networks by evolution? If yes,
which mechanism should be at work? We answer this question in the affirmative, by showing that
cooperation can be promoted by relying on simple strategies, i.e.
by imposing a limited set of rules which mimic the principles of
evolution, and let the systems evolve in time. 
It seems that network reciprocity is the
candidate mechanism for promotion of cooperation in some
networks. Indeed, if we describe wireless networks as graphs, an
analogy can be drawn with populations which are not well mixed.
The reason for this is that, given an energy constraint, one user
can interact only with the nodes which are in the range of its
transmission, thus forming a cluster of potential cooperators.

\begin{figure}[!htb]

\centering
\begin{tabular}{c c}
\subfloat[Initially, all nodes are defectors (in red). At first, node B transmits to node C directly. The energy required for direct transmission is a polynomial function of the distance $d_{BC}$.]{\includegraphics[scale=0.14]{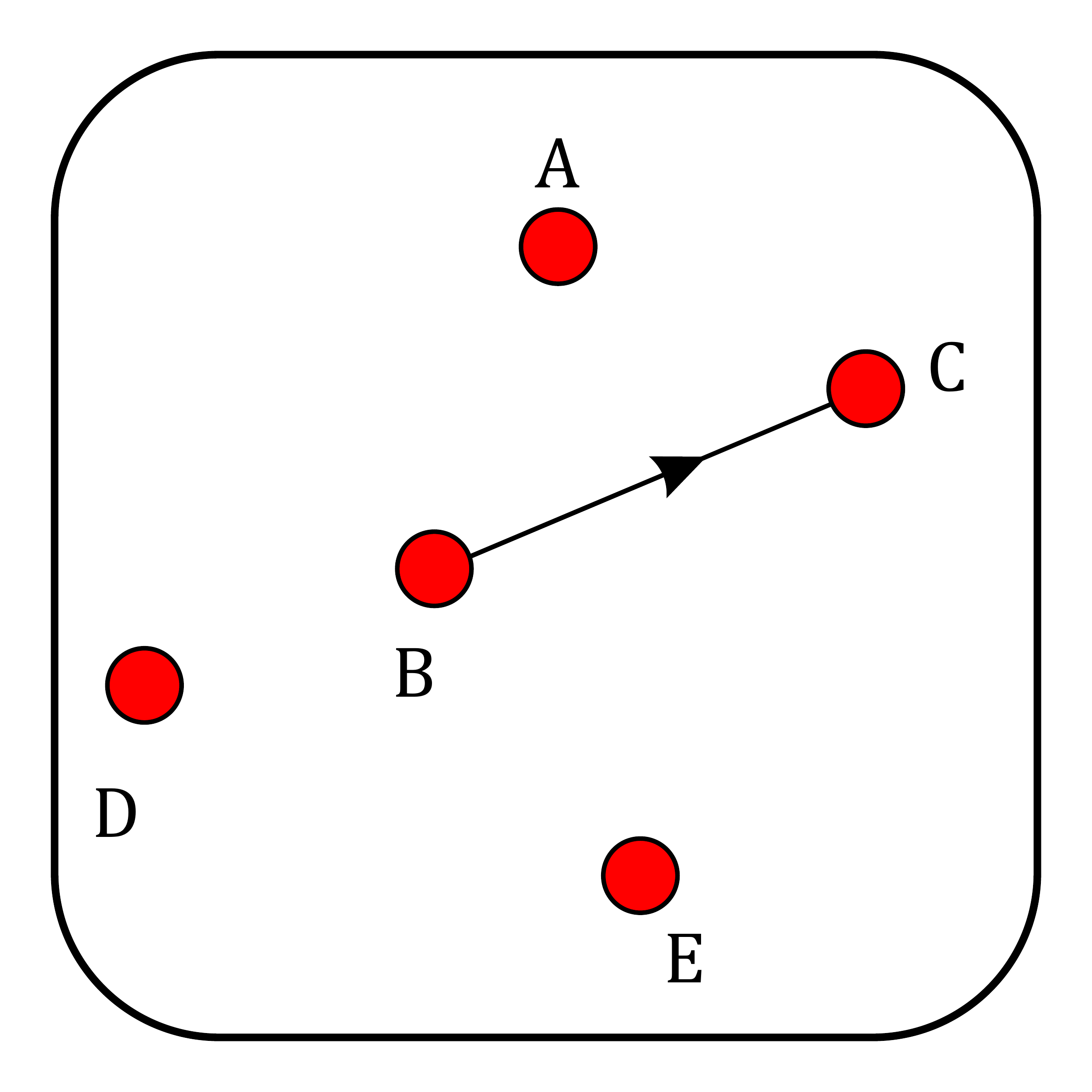}}
&
\subfloat[User A is randomly chosen to become cooperator, and assists B in the transmission. Since $d_{BA}<d_{BC}$, B can save energy by transmitting only to A.]{\includegraphics[scale=0.14]{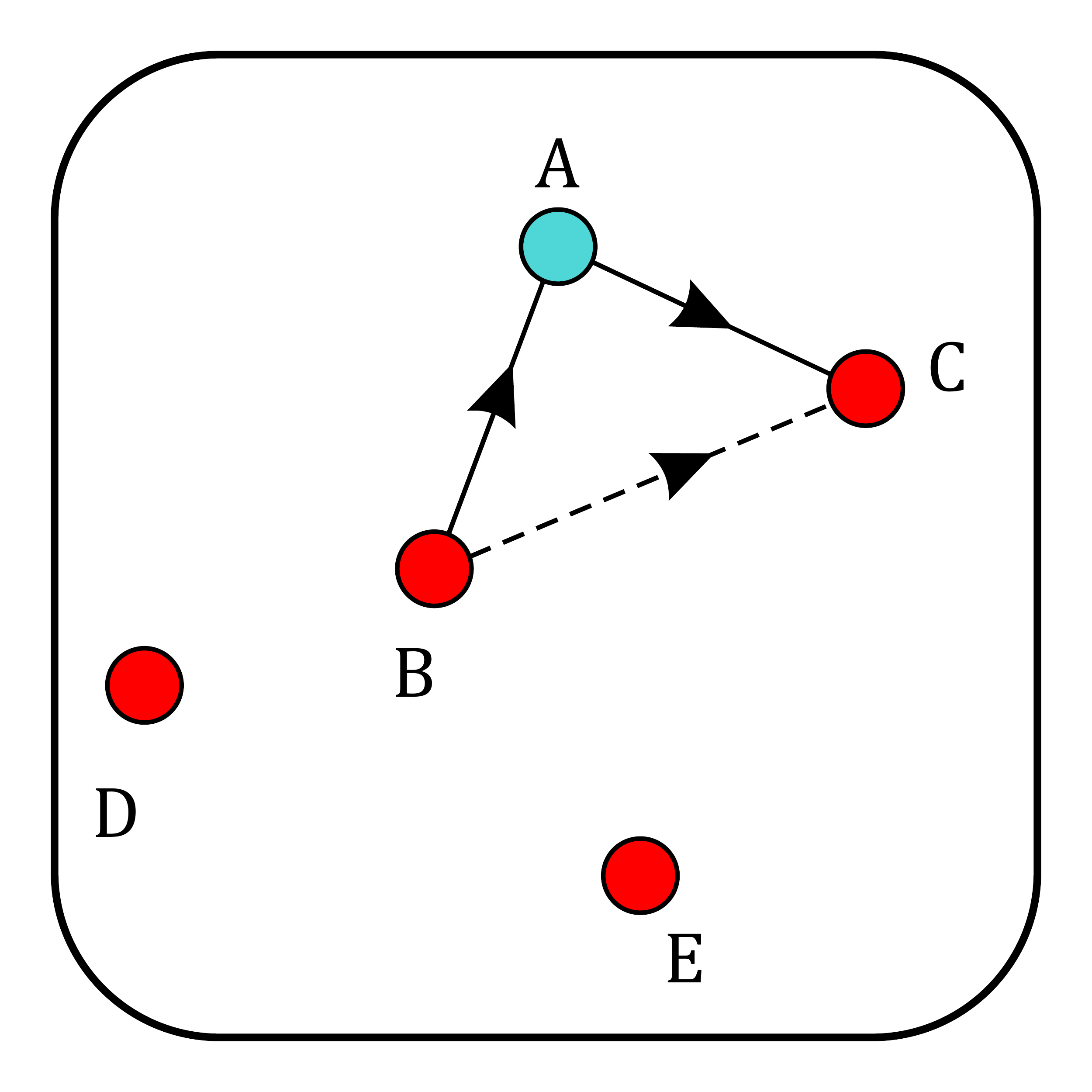}}

\\
\noindent

\subfloat[Since B observes increase of its fitness, it becomes cooperator and helps D in the transmission to A.]{\includegraphics[scale=0.14]{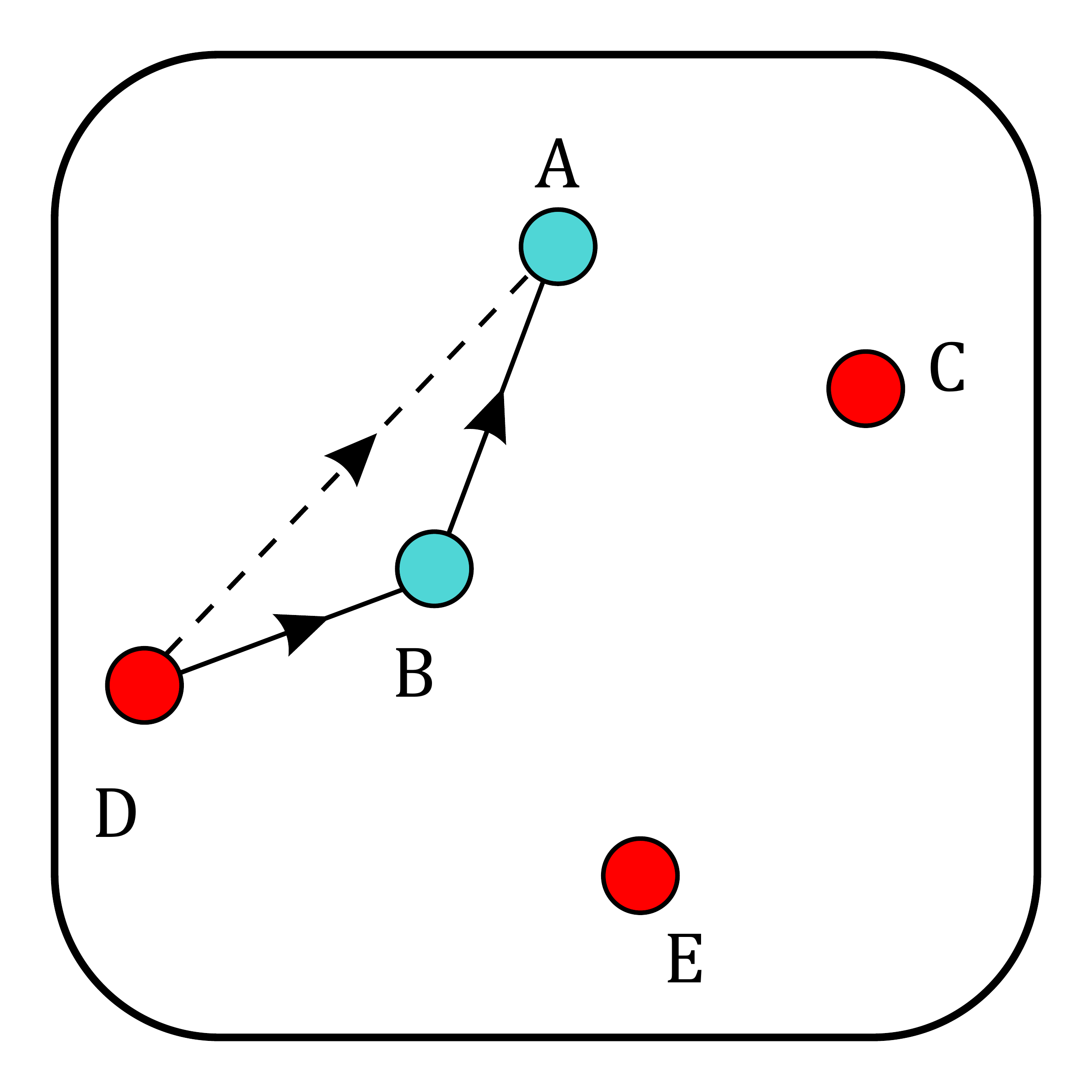}}
&
\subfloat[Finally, node A receives benefit for its initial cooperative act, as node B provides assistance for the wireless transmission to E.]{\includegraphics[scale=0.14]{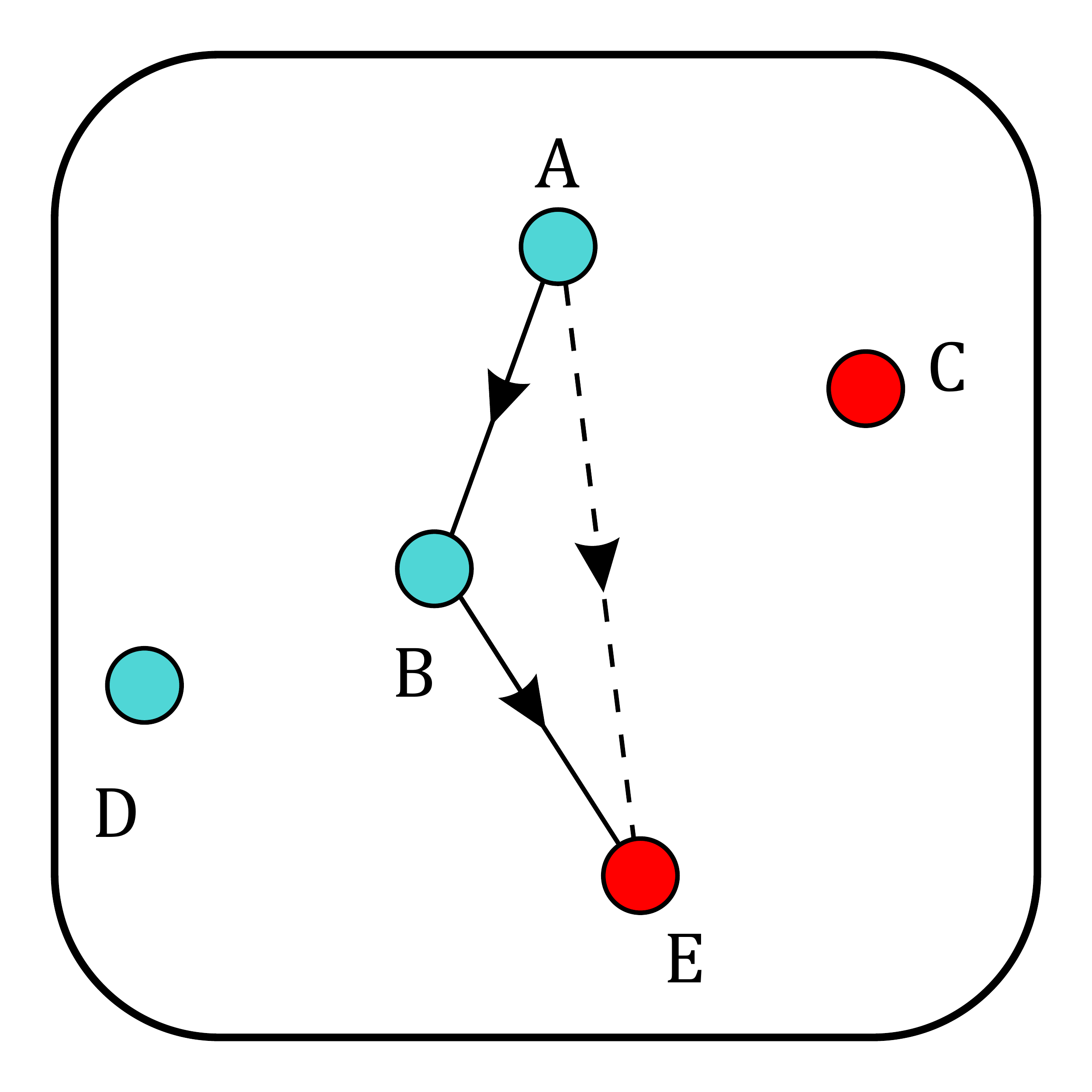}}
\end{tabular}

\caption{
Promotion of cooperation in wireless networks with selfish decision-making of the individual nodes. The mechanism behind can be seen as a form of network reciprocity, in the spirit of \cite{nowak2006}.
}

\label{fig:toy_example}       
\end{figure}
While we build on the legacies of communication protocols for
establishing cooperation in decentralized networks, our approach
differs in one important aspect. Namely, we do not assume
cooperation to be beneficial "by default", but we rather
adopt a game-theoretic approach where the network nodes
\textit{decide} whether to cooperate or not based only on their fitness, which is a quantity related to the
individual energy (power) consumption. We propose simple, decentralized strategies for the individual nodes and evaluate their energy efficiency.

Fig.~\ref{fig:toy_example} depicts a toy example which describes the mechanism for promotion of cooperation in wireless networks. The nodes in the network region of interest calculate their individual fitness (related to the energy they have on disposal) on a regular basis and adopt a simple strategy according to which they change from cooperators to defectors and vice versa. One example of such strategy is a form of TIT-FOR-TAT, where a node changes from defector to cooperator when it experiences benefit (increase of fitness) as result of the cooperative behavior of other node(s), and vice versa when there is no such benefit. As we can see, the introduction of a cooperator in the vicinity of a node involved in wireless transmission triggers cooperative behavior in that part of the network, even though the network nodes act in a selfish manner, i.~e. care only about their individual fitness. 
\section{Network Model}
\label{sec:network_model}
\subsection{Assumptions}
In order to investigate the emergence of cooperation in wireless
communication networks, we define a network
model which aims at capturing both the essence of wireless
communication networks and the graph models used in evolutionary
game theory. The model is such that it is still
rich enough to capture the essence of the energy consumption in
the network and the mechanism behind the emergence of cooperation,
yet simple enough to be able to interpret the observed phenomena.

We model the network as a graph where the users represent the
nodes and the edges are related to interactions between the nodes.
A time division multiple access (TDMA) approach is used where the
nodes take turns in transmitting their packets (no frequency
reuse). We divide the time scale in time slots of equal duration
and assume that one transmitter/receiver pair is activated at
random in each time slot. This multiple access scheme is known
to be
optimal \cite{zhao2005energy, el2006optimal, caire2004suboptimality}, at least in first approximation, from a minimum energy per bit perspective. Although this assumption simplifies the network analysis, it
may be regarded as restrictive, as in reality multiple
simultaneous transmissions can occur. Nevertheless, we expect that this simple scenario will be able
to capture the essence of the cooperative behaviour of the users, by performing
simulations over sufficient number of time slots, which averages users activity over time. We are tempted to
conclude that, from the perspective of the investigated phenomenon (emergence of cooperation), the simulation results will be
a reasonable indicator of the network behaviour in more general
scenarios.

We assume that the nodes are selfish in the sense that their 
objective is to be energy (power) efficient,
i.e to minimize the individual energy spent for 
transmission. As in game theory, we assume two types of nodes,
cooperators and defectors. 
The assumption of having
transmit packets of equal duration establishes
equivalence between power and energy and both are used
interchangeably throughout the paper.

We address network architectures with either direct (one-hop)
communication (in the case when there is no cooperator willing to
retransmit the packet), or two-hop communication (one
retransmission), in the case of presence of cooperator(s). This is
surely a simplification since, in general, information can also be
transmitted in a multi-hop fashion where information is relayed
from each source to its destination in successive transmissions
between intermediate nodes \footnote{The optimality of a
certain architecture (from perspective of capacity) depends on the
operating regime of the network (power-limited or interference
limited being two extreme examples). We refer the interested reader
 to \cite{ozgur2011operating} for the analysis of the capacity-scaling
performance of more general network architectures in different
operating regimes.}. Although being relatively simple, the
architectures we address are rich enough to describe the the network behaviour in
the game-theoretic framework. 

\subsection{The role of cooperation}
\subsubsection{Capturing the essence of the network behavior - A
fundamental network unit} 

\begin{figure}[!h]
\begin{center}
\includegraphics[scale=0.22]{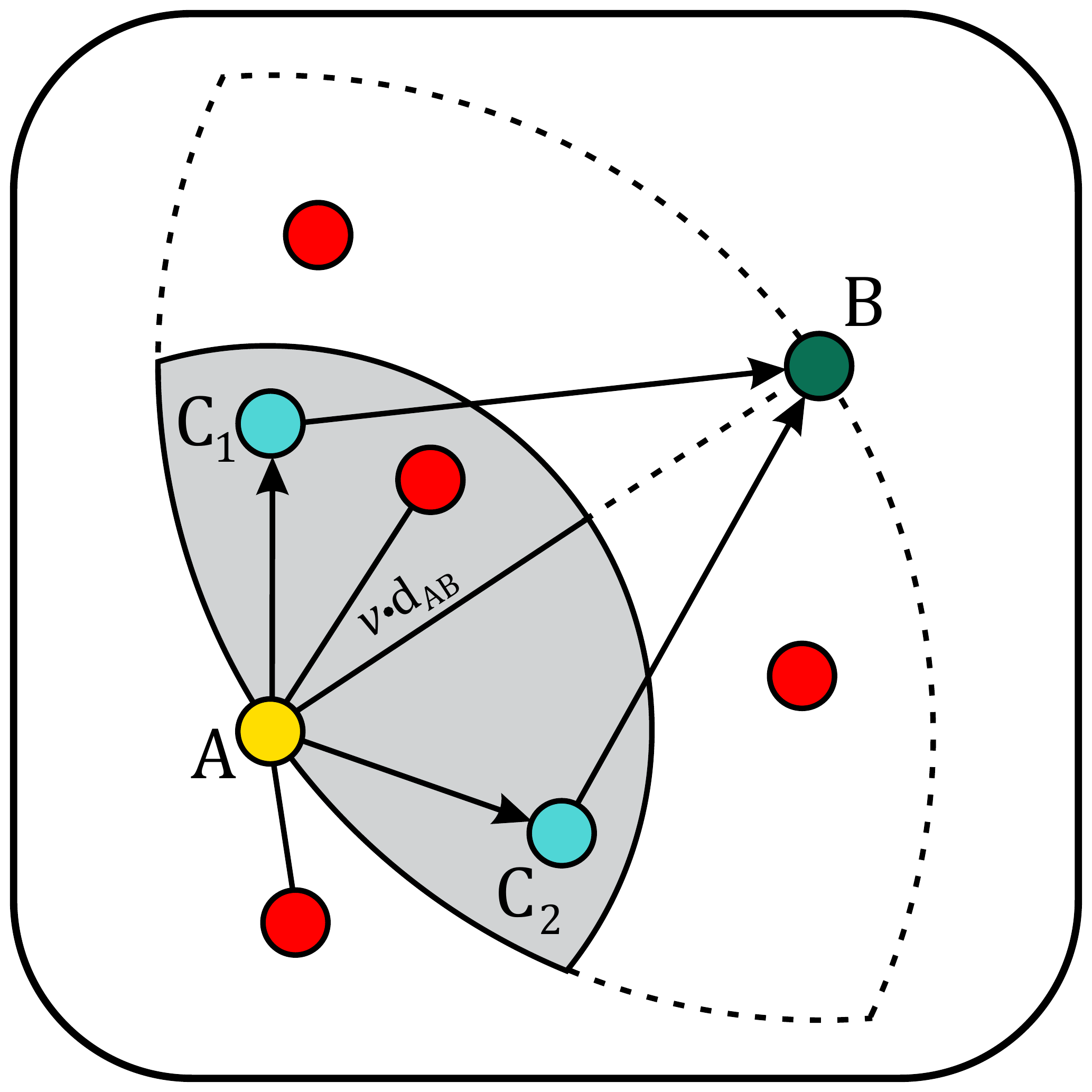}
\caption{Typical communication scenario: yellow circle -
transmitter ; green circle - receiver; blue circles - cooperators;
red circles - nodes that do not participate in the communication;
dashed line - the area where intermediate nodes are located; full
line - the area where the cooperators for the
transmission/receiver pair A/B are located; }
\label{fig:Intermediate_nodes}
\end{center}
\end{figure}

Let us look at a network snapshot which
is involved in one transmission (one time slot) between two nodes, A and B, as depicted in Fig.~\ref{fig:Intermediate_nodes}. Let
$P_{D}$ be the power A spends for direct transmission to B. As
result of the propagation effects, the received power at B is
$P_{R}=P_D/\left(K d_{AB}^\alpha \right)$, where $\alpha$ depends on the propagation characteristics of the
area (urban, suburban, rural, etc.) and $K$ is a propagation
constant. Typically, $\alpha$ takes values in the range
$2\leq\alpha\leq 4$.. We define the
signal-to-noise ratio at the receiver as $SNR=P_{R}/\sigma^2$,
where $\sigma^2$ is the noise variance. We say that the
transmission is \textit{successful} if the receive SNR exceeds a certain threshold required for reliable reception, $SNR\geq
SNR_0=P_{R_0}/\sigma^2$, i.~e. A should transmit with power $P_{D}\geq Kd_{AB}^\alpha
P_{R_0}$. We assume perfect
power adaptation and take
that node A adjusts the transmit power to the distance
$d_{AB}$, such that it meets the receive SNR requirement
\textbf{exactly}\footnote{This is a simplification since for
this adaptation to work, A should know the network topology (the
distance to B) or to have feedback from B
about the receive SNR, in order to adjust the transmit power.}.

We say that a node $C$ is in \textit{range of} A, or
\textit{connected} to A, if it can "hear" A's transmission to B.
Under the assumed power adaptation, a node C is in the range of A
if $d_{AC}\leq d_{AB}$. Additionally, we assume that C is a
potential cooperator only if $d_{CB}\leq d_{AB}$. This assumption
is reasonable, since otherwise the cost of retransmission would be
higher than the cost of direct transmission \footnote{In general,
one could allow for a cooperator C to be found at a distance
$d_{CB}\geq d_{AB}$. In this case cooperation might still be
beneficial, if the cost of cooperation is shared
between several cooperators. This 
assumption could influence the network behaviour, particularly in the case of the network architecture with
a central infrastructure node, addressed in Section
\ref{sec:experiment_setup}.}. 
According to the scenario we address, the region containing
potential cooperators is described by
\begin{equation}
\label{eq:intermediate}
  d_{AC} < d_{AB};\:\:d_{CB} < d_{AB}.
\end{equation}
A node C which fulfills (\ref{eq:intermediate}) is called
\textit{intermediate} node.
As depicted in
Fig.~\ref{fig:Intermediate_nodes}, potential cooperators are
located in the area enclosed with a dashed line. This constellation represents a typical scenario which captures the essence of the
communication in the network and can be thus regarded as
\textit{fundamental network unit}. If an intermediate node C decides to help A in the
transmission, it will retransmit the signal
received from A to B.
The benefit that the node A obtains from the cooperative act of C
is that, in the presence of the cooperator C, A can decrease the
transmit power to a value lower then the power required for direct
transmission, $P_{I}\leq P_{D}$, 
which defines the benefit of the
cooperative act as $b=P_{D}-P_{I}$.

In general, for a given transmitter/receiver pair A/B there can be
more then one cooperator. In this case they can either share the
cost for cooperation or let one cooperator pay the overall cost.
For simplicity, we only address the case where only the closest
cooperator to B retransmits the signal and leave different
approaches to cost sharing for future study. At first sight, this
simplification might seem suboptimal, since in this way in one
time instant the cost of cooperation is paid by a single
cooperator. However, this effect will be averaged over time since
for different transmitter/receiver pairs, the choice will fall on
different cooperators (with high probability).

\subsubsection{Protocols for establishing cooperation}
In order for cooperation to work, the involved nodes should
exchange some kind of control messages, i.~e. they need to
establish a \textit{cooperation protocol}. We hereby identify two
protocol scenarios. For both protocols, we implicitly assume that
the nodes have the amount of knowledge of the network topology
required for both protocols to work.
\paragraph*{Cooperation protocol 1}This cooperation protocol
assumes that user A sends a low-rate \textit{request to relay}
message within a region of radius $\nu \cdot d_{AB}$, where $0\leq
\nu \leq 1$. The parameter $\nu$ has the role to further reduce
the range of A's transmission. 
With this convention, instead of
engaging all cooperators from the area defined with
(\ref{eq:intermediate}) in the retransmissions, we account only for
those located in the area enclosed by the full line in
Fig.~\ref{fig:Intermediate_nodes}. By introducing the parameter $\nu$, one avoids sending
\textit{request to relay} messages at large distances.
Additionally, it is expected that the cooperative act is most
beneficial for the transmitter A when the potential cooperators
are found (relatively) near to the transmitter, since this lowers
the cost of direct transmission. Although this choice could
potentially increase the total cost of cooperation, this will not
affect significantly the cost that individual cooperators pay.
Indeed, the total cost of cooperation is
distributed between the cooperators in the region of interest,
either directly (by having more cooperators assisting a single
transmitter), or indirectly (by choosing different cooperators
over time). 

Having received the \textit{request to relay} message (which
besides the sender A, also identifies the destination B),
cooperator nodes located in this region will send back an
acknowledgement to A that they accept to relay (retransmit). If
more than one cooperator is present, A decides which of the
cooperator(s) will actually retransmit. The subtle characteristics
of this protocol is that node A can be selfish in the decision
process in the sense that it can choose only one cooperator-the
nearest one, in order to minimize its own transmit power. In that
case, in the final phase of the transmission node A will transmit
a signal with power adopted to the distance of the nearest
cooperator, which will then retransmit the message. We note that
this behaviour increases the cost of cooperation and is also
suboptimal from the point of view of the total energy consumption
of the network.
\paragraph*{Cooperation protocol 2}
In the first phase of this cooperation protocol user A also
transmits a \textit{request to relay} message in the region of
radius $\nu \cdot d_{AB}$. In the second phase, the cooperators
coordinate among themselves and one of the cooperators sends node
A a general confirmation that \textit{there are cooperators}
present in this region (without disclosing their identity). In the
case of no response, A assumes that there are no cooperators. If there is a confirmation of
cooperators present, in the next phase A will transmit the message
with reduced power $P_I=\nu^\alpha P_D$. Having received the
signal from A, it is now on the cooperators to decide which one(s)
will retransmit the message. This can be done by a form of
coordination between the cooperators\footnote{One such form of coordination is \textit{quorum sensing} which refers to the phenomenon in which the accumulation of signalling molecules in the surrounding environment enables a single cell to assess the cell density so that the population as a whole can make a coordinated response.}. 
If one cooperator is chosen to retransmit the
message, the natural choice falls on the one which is closest to
the destination B.
The difference with the first protocol is that now the
coordination is performed among "trusted nodes", i.~e.
cooperators. While in the first protocol the transmit node A
(which can be a defector in general) dictates the cooperation, in
the second one it is the cooperators who decide on the details of
the cooperative act\footnote{Similar behavior has been observed
while studying structured populations \cite{nowak2006}. In these
populations, once introduced by chance, cooperation persists by
the formation of clusters of cooperators which help each other.}.

In the scenarios we address, we adopt the second cooperation
protocol, as we expect that it would favor cooperative
behavior. The motivation behind is that the adoption of the first cooperation protocol would
encourage the transmitter node to save energy by transmitting only
to the nearest of the potential cooperators, which would
eventually undermine the efforts of the cooperative nodes and
discourage spreading of cooperative behavior in the
networks\footnote{This conjecture has yet to be supported by
simulations.}. On the other hand, the second protocol leaves less
room to the transmitter nodes (which are in general not
trustworthy) to misbehave, and allows for arrangements among the
trustworthy, cooperative nodes.
\section{A Game-theoretical Framework for Energy-efficiency Analysis in Wireless Networks}
\label{sec:game_theory} Game-theoretic approach to modelling of
phenomena assumes existence of some quantity -- utility, or
benefit -- that
 units in the system try to maximize \cite{lasaulce2011game}. In some scenarios the agents may choose to help the others, i.e. to cooperate -- this is modeled
 by the cost they pay for the cooperation. Some agents choose their strategy to be selfish, i.e. they defect, and thus avoid any costs.
 The cost of cooperative act implies that the cooperators would have smaller fitness than the defectors, i.~e. that
 natural selection of the fittest would favors defectors. However, recent results in evolutionary biology suggest that
cooperation can be favored by natural selection, if a certain
mechanism is at work. Indeed, there are observations and theoretic
 analyzes of cases when cooperation persists -- there is at least a fraction of cooperators present in the
 population.

\subsection{Definition of Fitness in Wireless Networks} Following the analogy with evolutionary
biology, we will define fitness of the individual network nodes. 
Intuitively, the fitness has to be related with the energy
consumption of the individual nodes. Ideally, the appropriate
fitness function has to be simple enough to be evaluated locally
(possibly without requiring complex processing and memory). On the
other hand, it has to be rich enough to capture the essence of wireless transmission and network dynamics. 
We will define two discrete time scales,
according to which the fitness will be evaluated.
The fitness function is evaluated at the end of a block of
duration $T$ slots. The network performance is observed over $N$
such blocks (iterations). Since we have two time scales, we introduce two indices, $t$ and $n$,
where $t\in\left\{0,1,\ldots,T\right\}$ indicates the time slot
and $n\in\left\{0,\ldots,N\right\}$ indicates the iteration. The
fitness is then a function of $n$ and $t$, $F=F(n,t)$. We note
that, in order to be consistent with the definition of $t$, we
denote the initial iteration (of duration $T$) as $0$-th
iteration. Additionally, we denote the initial fitness as
$F(0,0)=F_0$. Now, let us define
\begin{equation}
\Delta f(n,t)=F(n,t)-F(n,t-1),\:\: t=1,\ldots,T ,
\end{equation}
which measures the difference in the fitness evaluated at two
consecutive time instants $t-1$ and $t$, of the $n$-th iteration.
In our model, $\Delta f(n,t)$ for the network node C is
defined as
\begin{equation}
\Delta f(n,t)=-\alpha \left( 1 - \beta \right) \left[ P_D-P_I \right] - \gamma \delta P_C(J)
\end{equation}
where $\alpha, \beta, \gamma, \delta \in{\lbrace0,1\rbrace}$ are parameters
which indicate packet transmission and presence of cooperators and
defectors. In particular, $\alpha=1$ when C has a packet to
transmit; $\beta=1$ when C has at least one cooperator as a
neighbor; $\gamma=1$ when C is connected to at least one active
node at that time instant; and $\delta=1$ corresponds to C being a
cooperator (otherwise the parameter values are zeros). We note that the above parameters are also functions
of $n$ and $t$. However, whenever there is no ambiguity, and in
order to simplify the notation, we will skip these indices. Having introduced $\Delta f(n,t)$, we can define the fitness of
the node C in the following way
\begin{align}
\label{eq:fitness}
F(0,0)&=F_0,\nonumber\\
F(n,t)&=F(n,t-1)+\Delta f(n,t),\nonumber\\
F(n+1,0)&=F(n,T)
\end{align}
where $n=0,1,\ldots, N-1$ and $t=0,1,\ldots, T$. Defined in this way, the fitness reflects both the \textit{energy
saving} of the individual nodes when their transmissions are
assisted by cooperators, e.~g. the benefit they receive, and their
energy expenditure when they assist other nodes in the
transmission, i.~e. the cost they pay for cooperation. Additionally, we define the quantity
\begin{equation}
\Delta F(n)=F(n,T)-F(n-1,T),\:\: n=1,\ldots,N
\label{eq:delta_fitness}
\end{equation}
to be the change of fitness between two consecutive iterations.
\subsection{Game-theoretic Strategies}
In this work we study four different strategies of cooperative
behavior in wireless networks. In the present approach we assume that all network nodes adopt
the same strategy during the simulations. This approach certainly
does not cover some more general scenarios, for example the one
when the individual nodes are able to choose their strategy at
random, or according to some rule. Nevertheless, we expect that
the results from our analysis will fairly well indicate the
general trend and, as such, will be useful in the evaluation of
the fundamental limits on energy efficiency in decentralized
networks.

The first strategy addresses the trivial case when there is no
cooperation between the nodes, i.~e. all nodes are defectors. We
denote this strategy by DEF. The second strategy addresses the
case where all nodes cooperate and will be denoted as COOP. It
corresponds to a scenario where cooperation is "
enforced" in the network. 
Besides these "trivial" cases, we will concentrate on
be strategies which where the
individual nodes are essentially selfish and decide whether to
cooperate or not based solely on their individual fitness. In the
scenario that we propose, at the end of the $n$-th iteration the network
nodes make a simple decision whether to cooperate or defect in the next iteration,
based on the change in the fitness $\Delta F(n)=F(n,T)-F(n-1,T)$,
as defined in (\ref{eq:delta_fitness}).

We will distinguish between two simple and intuitive strategies
for this scenario. According to the first one, if the node
observes an increase in the fitness, $\Delta F(n)>0$, it will
retain the previous status in the next iteration. If, on the other
hand, the node observes decrease in the fitness $\Delta F(n)<0$,
the node will change its behavior, i.~e. a cooperator will become
a defector and vice versa. We observe that from the perspective of
a single node, the game resembles the repeated prisoner's dilemma
\cite{axelrod1981evolution}. In this regard, the above described
strategy corresponds to the well known \emph{win-stay, lose-shift} (WSLS)
and is based on the simple idea of retaining the previous status
when the node is doing well, but changing otherwise. 
According to the other strategy for this scenario, the node will
decide to cooperate in the next iteration if it observes an
increase in the fitness, $\Delta F(n)>0$. Otherwise, it will
defect. According to this strategy, a defector will become
cooperator and a cooperator will stay cooperator, if $\Delta
F(n)>0$. Otherwise, the node will choose to defect. We note that
the increase in fitness reflects the average behavior of the
adjacent nodes, in the sense that the reason for the fitness
increase is the cooperative behavior of some of the adjacent
nodes. In the context of the repeated prisoner's dilemma, this
strategy resembles the \emph{tit-for-tat} (TFT) strategy which is based
on the idea of mimicking the other node(s) behavior in the
previous turn. This means that the node will become cooperator
only if it observes cooperative behavior of other nodes which is
reflected in the increase in the fitness. 
\section{Examples of Network Architectures: Description of the Experiments}
\label{sec:experiment_setup}
\subsection{Network Architectures}
\label{sec:network_architectures}

\begin{figure}[!htb]
\centering
\begin{tabular}{c c}
\subfloat[]{\includegraphics[scale=0.13]{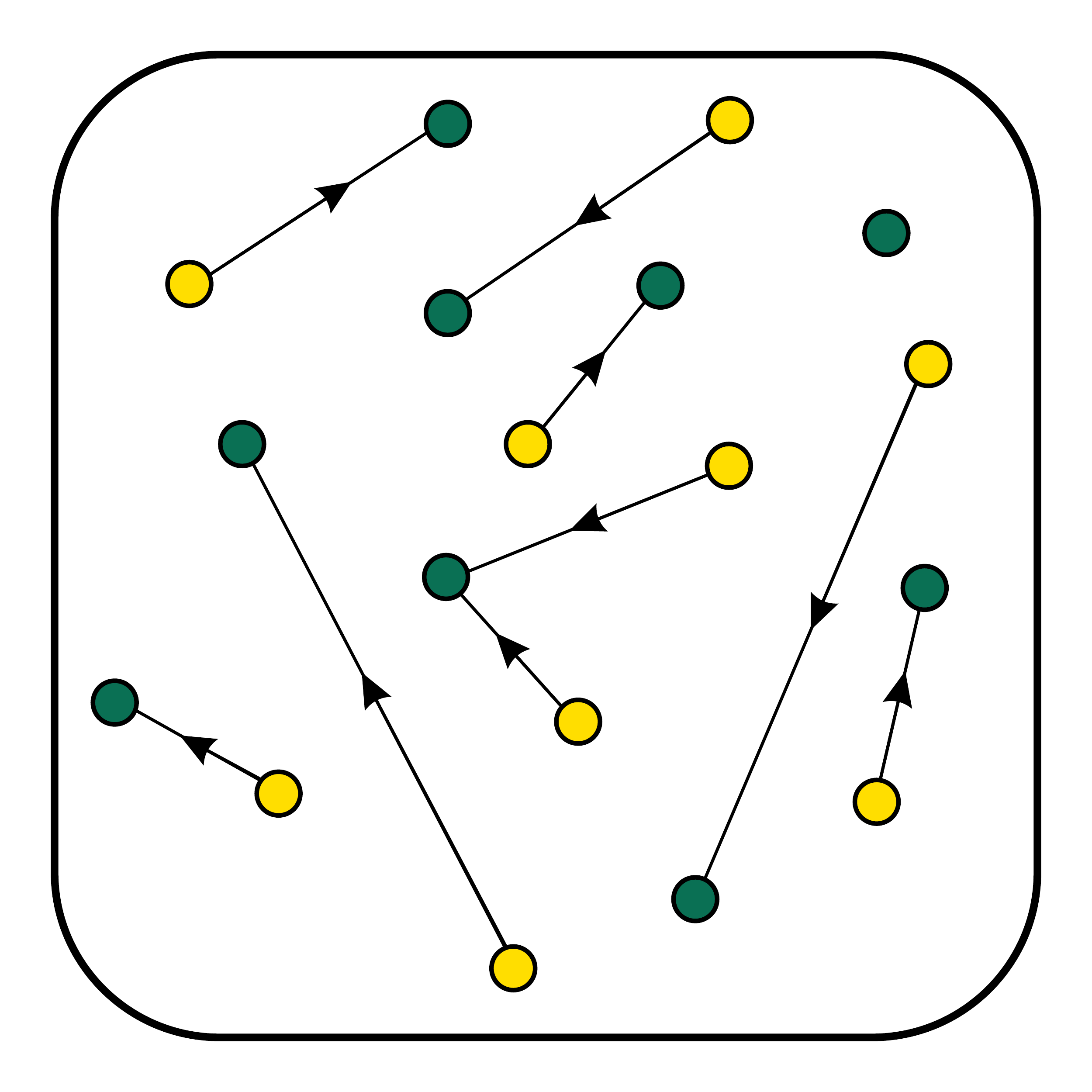}}
&
\subfloat[]{\includegraphics[scale=0.13]{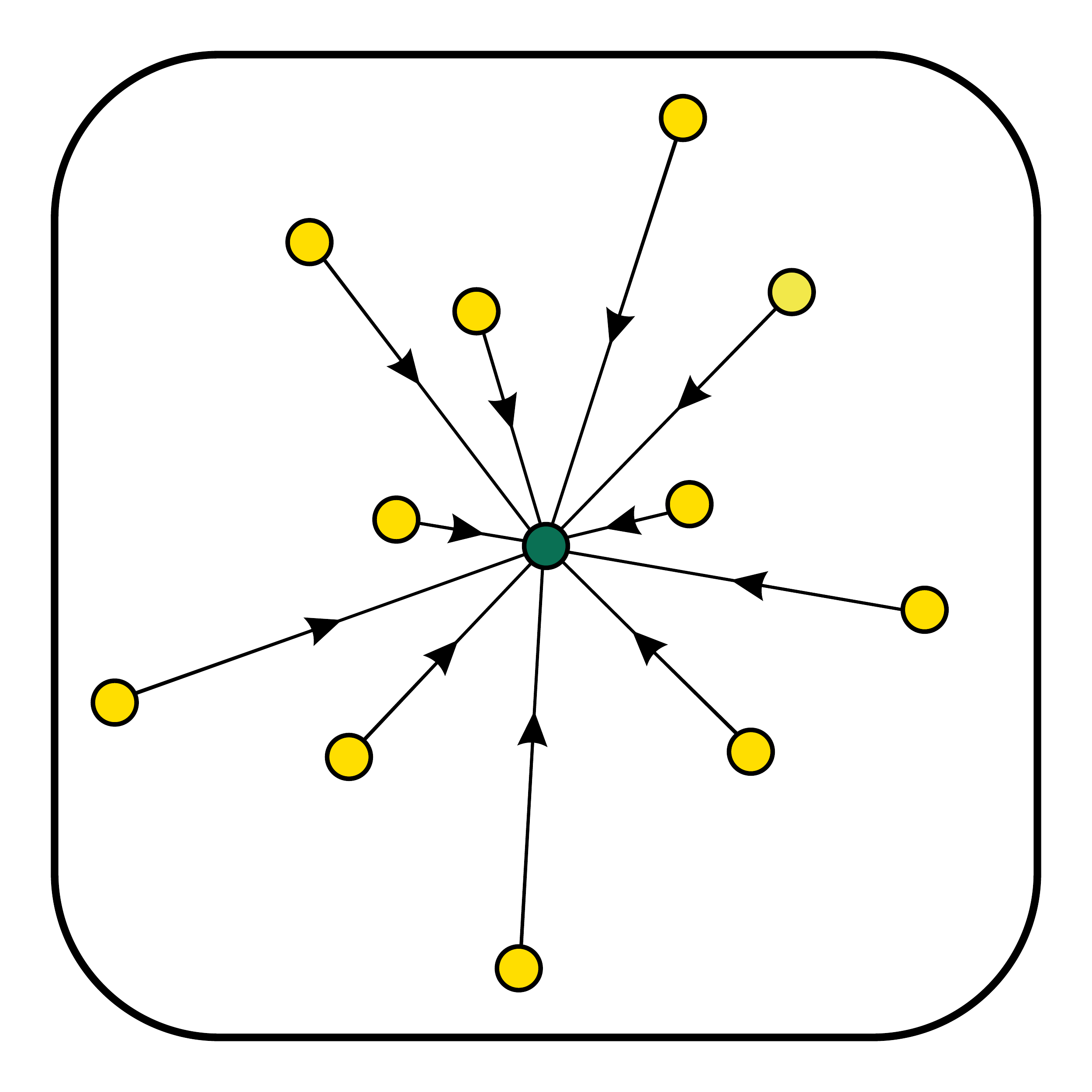}}
\end{tabular}
\caption{Network architectures where cooperative behavior is considered: a) wireless ad hoc network; b) network with central infrastructure node.}
\label{fig:architectures}       
\end{figure}

We investigate the cooperative
behavior in two wireless network architectures, \textit{wireless
ad hoc network} and \textit{network with a central infrastructure
node}. A graphical representation for the two network architectures is given in Fig.~\ref{fig:architectures}.
\paragraph*{Wireless ad hoc network} A wireless ad hoc network represents a collection of wireless
nodes that self-configure to form a network without the aid of any
established infrastructure. Immediate applications of ad hoc
networks include emergency and battlefield networks, metropolitan
mesh networks for broadband Internet access, and sensor networks.
In addition to these pending applications, ad hoc networks are
closely related to the science of networks in other fields,
including biology, economics, and air and automobile
transportation.
\paragraph*{Network with central infrastructure node}
The second architecture we address is asymmetric and assumes that
the users in one area (circle for example) transmit their signals
to a central infrastructure node, e.~g. access point, relay, or
base station. This architecture is particularly relevant in the
context of the emerging trends in the design of future wireless
networks, which rely on \textit{dense} and \textit{heterogeneous
deployment} of the wireless network infrastructure. Dense
deployment means pairing traditional macrocells with pico or
femtocells in densely populated urban environments; heterogeneous
deployment means combining the current cellular network
infrastructure with a parallel offloading infrastructure, based on
multiple radio-access technologies (such as GSM, UMTS, WiFi, and
device-to-device communication) \cite{zhao2009scalability}.
The simulation set up for both network architectures is as
follows. We place $M$ wireless nodes uniformly at random in a circle of
 unit radius $r$ (we are
interested only in the relative performance of the different
strategies and not in the absolute value of the consumed energy).
The nodes send their messages during time slots of fixed duration
(same for all nodes), where in each time slot exactly one
transmitter/receiver pair is activated at random. We group the
time slots in a block of length $T$, which denotes one iteration. The network behaviour is observed over
$N$ iterations. Additionally, $N_t$ different network configurations are tested. The parameters selected for our simulations are: $M=30$, 
$\alpha = 4$, $T = 1000$, $N = 1000$ and $N_t=1000$.

Assuming the aforementioned strategies for the behavior of the
individual users, the aim of the analysis/simulations will be to 
investigate the cooperative behaviour in the network and to  
evaluate the performance of the different strategies
in terms of individual and global energy consumption, for both
network architectures. As we will see, although both architectures
share certain similarities, the adoption of the same game-theoretic
strategy may yield qualitatively different network behavior. 

While simulating the performance of TFT and WSLS, we will 
assume that in the initial iteration all users are
defectors. At the end of the initial iteration we choose one user
at random to become cooperator. 
According to the change of the fitness defined in
(\ref{eq:delta_fitness}), the users determine their behaviour
during the next iteration (cooperate or defect) according to the
TFT or the WSLS strategy. The performance will be compared with
COOP (all cooperators) and DEF (all defectors), which serve as upper respectively 
lower bound for the performance. 
\subsection{Simulations: Wireless ad hoc network}
In this set up we assume peer-to-peer communication where each transmitter/receiver pair is equally
likely.

\begin{figure}[!htb]
\centering
\begin{tabular}{ @{\hspace{0pt}} c @{\hspace{-3pt}} c @{\hspace{-3pt}}  c @{\hspace{-3pt}}  c @{\hspace{0pt}}}
  {\includegraphics[scale=0.09]{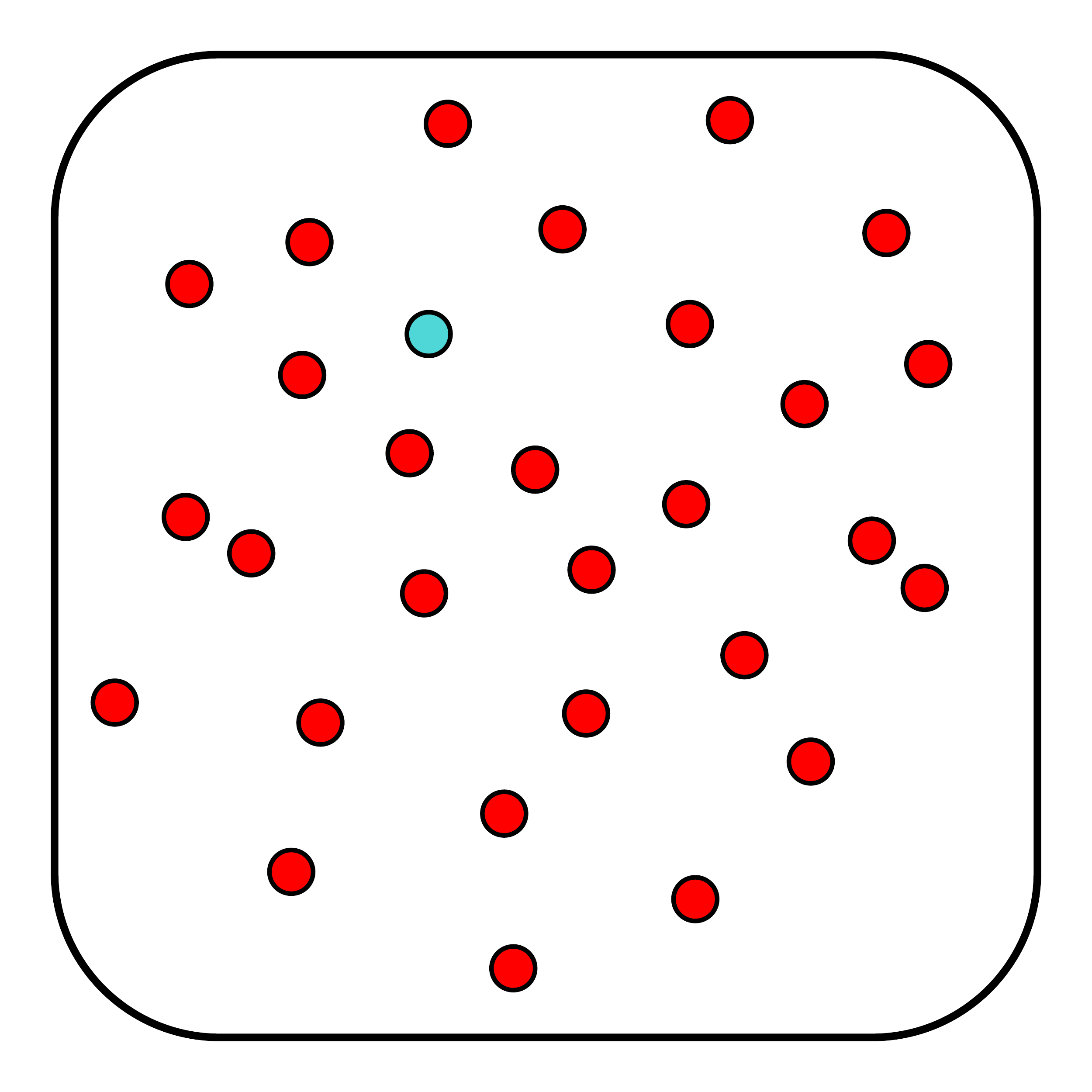}}
& {\includegraphics[scale=0.09]{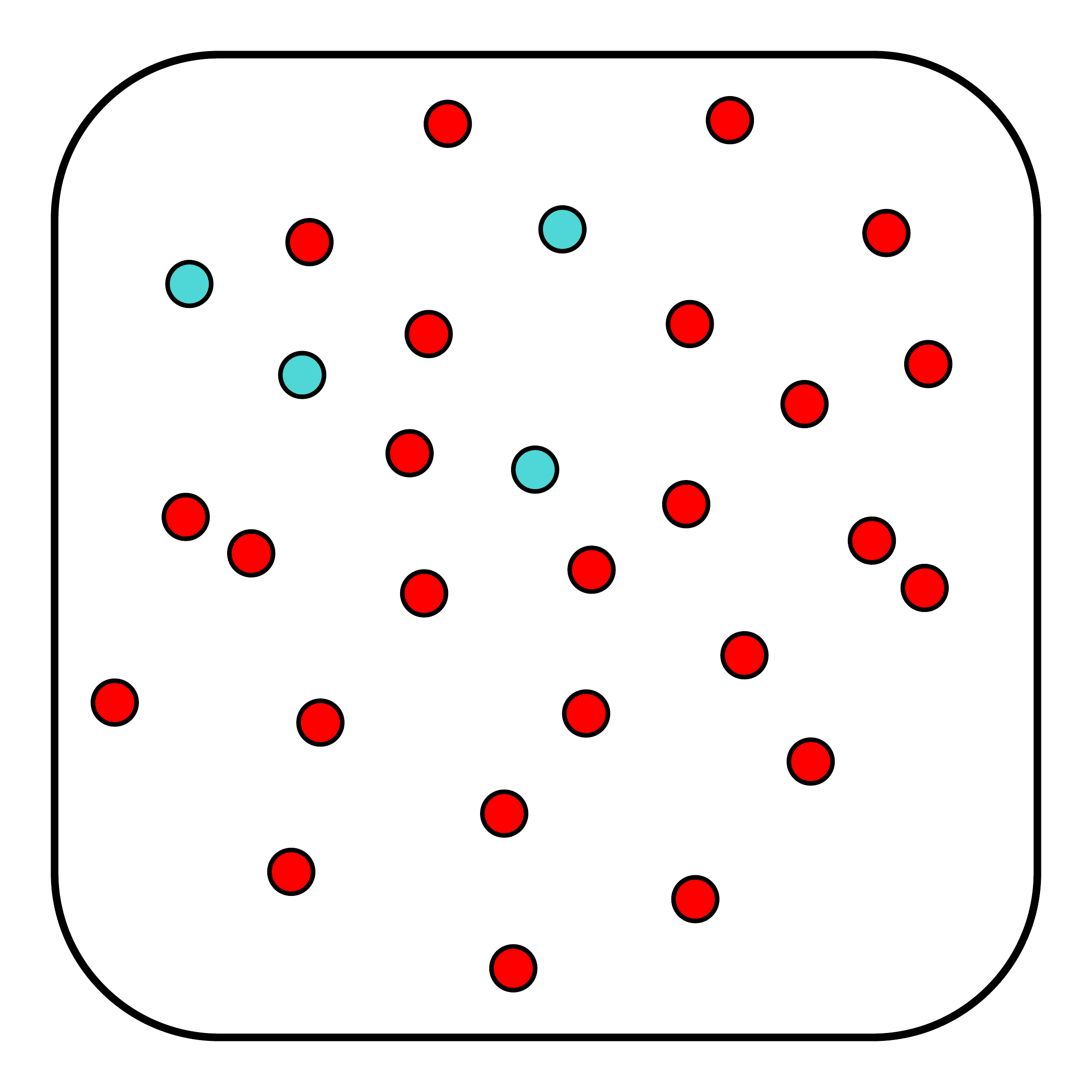}}
& {\includegraphics[scale=0.09]{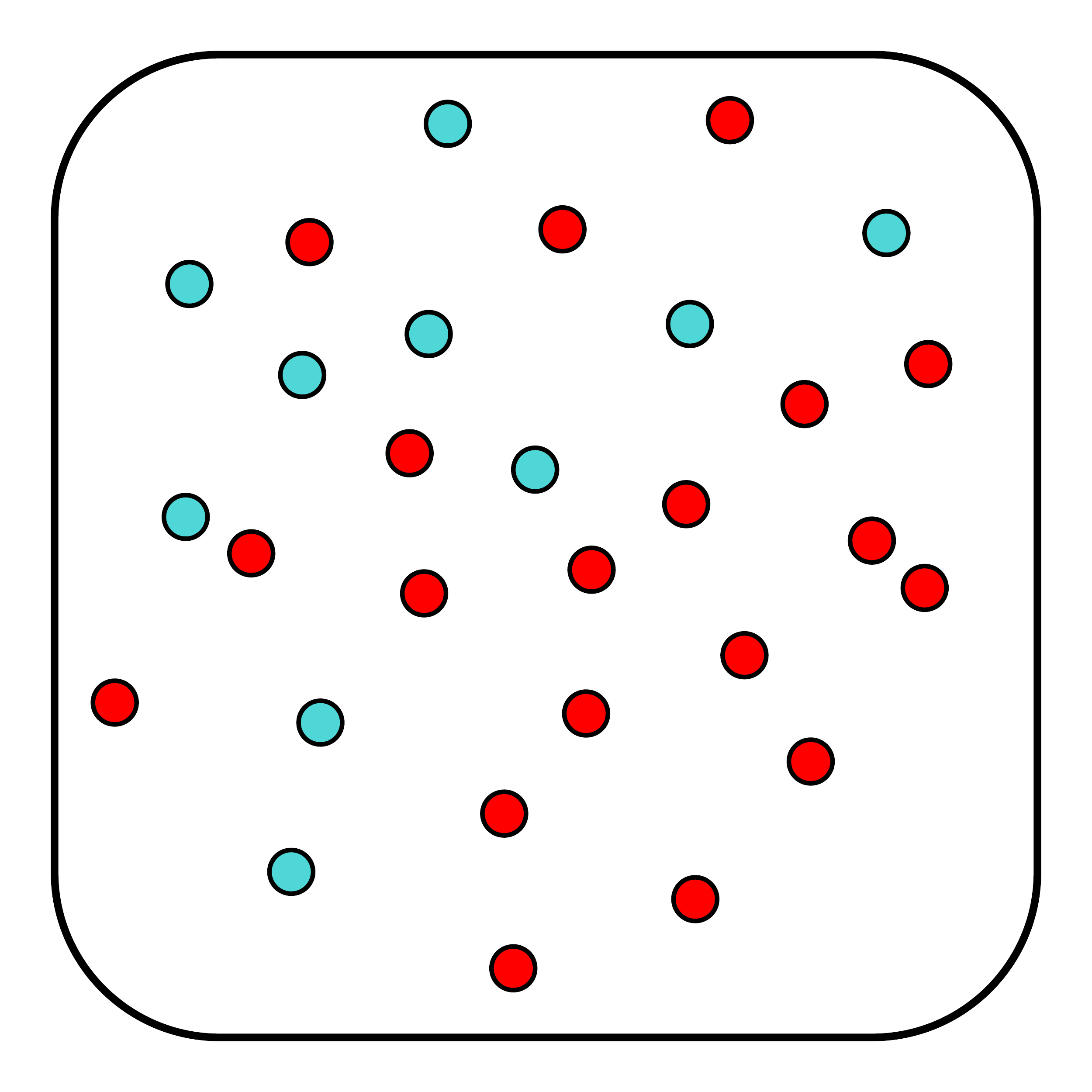}}
& {\includegraphics[scale=0.09]{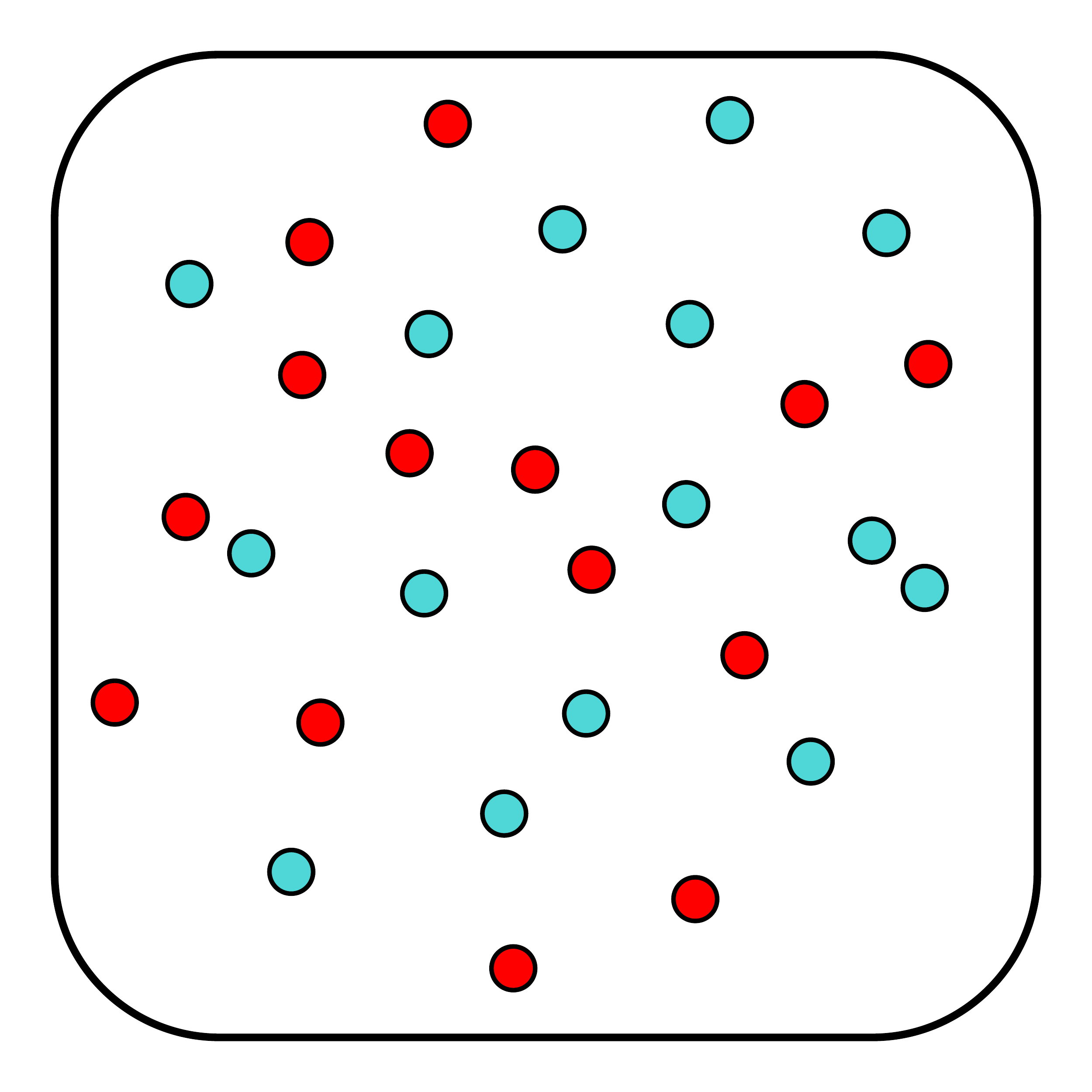}}
\end{tabular}
\caption{Emergence of cooperation in wireless ad hoc networks.
Red circles--defectors; blue circles--cooperators.
After the introduction of a single cooperator, cooperative
behavior spreads in the network and persists over time.}
\label{fig:spreading_of_cooperation}       
\end{figure}

\textit{1) Cooperative behaviour:}
In the case of TFT and WSLS, simulation results indicate that, once a single cooperator is
introduced in the network (by chance), the network evolves such
that cooperation spreads through the network, even though the decision 
making of the individual nodes is based on
a (rather simple) evaluation of the individual fitness. Fig.~\ref{fig:spreading_of_cooperation} depicts the effect of spreading of cooperation over time. In addition, 
the simulation results show that the emergence and stability of
cooperation is fairly robust to the random placement of the
initial cooperator. This, as we believe, is mainly due to the
assumption that each transmitter/receiver pair is equally likely,
which brings symmetry to the problem. We note that the effect of
spreading of cooperation, in general, also depends on the activity
of the users and the duration (length) of one iteration. We recall
that in our scenario we assume that in each time slot (at least)
one transmission takes place in the network, and that one
iteration involves a long number of time slots. This means that
during the first iteration there will almost certainly
\textit{exist} network node(s) which will benefit from the
cooperative act of the initial cooperator, which is exactly the
\textit{prerequisite} for cooperation to spread in the network. If
this is not the case, cooperation is expected to vanish in the next
iteration.

\textit{ 2) Energy consumption:} Before proceeding,
we first evaluate the influence of the choice of the parameter
$\nu$, defined in Section \ref{sec:network_model}. 
The simulations show that the value of $\nu$ for which the total energy consumption is
minimal varies slightly with the choice of strategy, but can be located close to
$0.39$. This value is used in the rest of the simulations.

When the users adopt the DEF strategy, a transmitter A
communicates with a receiver B by direct transmission. In the case
when the users adopt the COOP strategy, for a transmitter/receiver
pair A/B, user A receives benefit from the cooperators located in
the area defined by (\ref{eq:intermediate}), reflected in the fact
that it can
decrease the transmit power. The simulations show that when all nodes cooperate the total (average) energy
consumption is reduced by 60\% as opposed to the case
when all nodes defect, as indicated in Table~\ref{tab:mean_std}, which illustrates the standard deviation
of the individual energy consumption as function of
the network geography (distance from the center). 
\begin{table}[!h]
\caption{Mean and standard deviation of the energy $\mathcal{E}$}
\label{tab:mean_std}
\begin{center}
\begin{tabular}{ l | r | r }
  strategy & mean($\mathcal{E}$) & std($\mathcal{E}$) \\
  \hline
   DEF  & 1.00000 & 0.72150 \\
   COOP & 0.39755 & 0.13093 \\
   TFT  & 0.48858 & 0.11559 \\
   WSLS & 0.60966 & 0.01671 \\
\end{tabular}
\end{center}
\end{table}

\begin{figure}[!h]
\begin{center}
\includegraphics[scale=0.39]{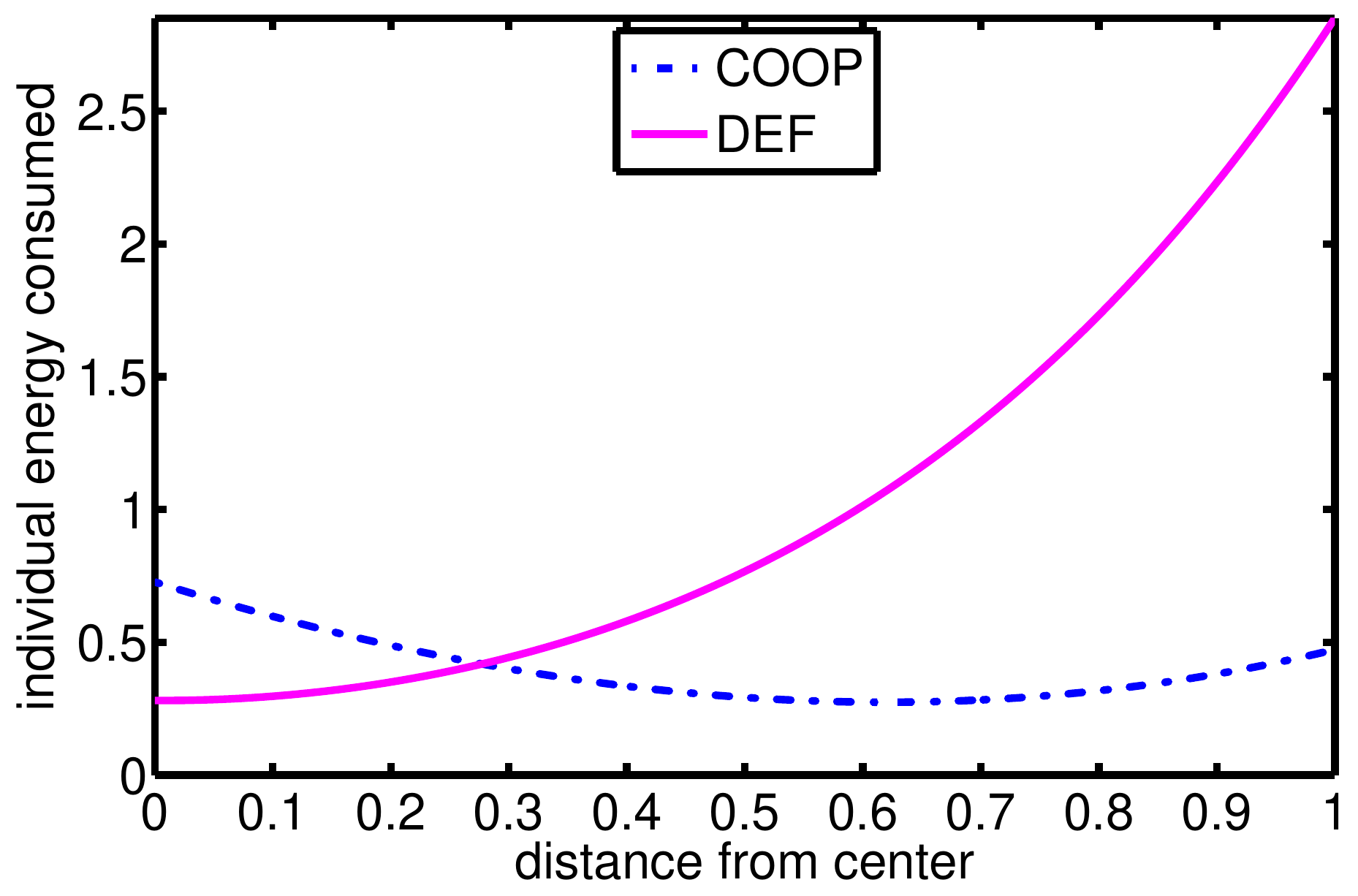}
\end{center}
\caption{Individual energy consumption depending on the location.}
\label{fig:energy_spent_vs_r_def_coop}
\end{figure}

Apart from the
reduction in average energy consumption, cooperation leads to a
more-fair energy consumption among the individual nodes. Indeed,
when there is no cooperation the nodes which are located further
from the center are at a disadvantage as the average distance to
the rest of the nodes is larger compared to the nodes which are
located near the center, leading to increased energy consumption
for transmission. The introduction of cooperation lessens this
imbalance to some extent. An illustration of this effect is shown
in Fig.~\ref{fig:energy_spent_vs_r_def_coop}. As we can see, the
introduction of cooperation balances the amount of energy spent by
the individual nodes, and decreases the effect of the network
topology on the individual energy consumption.

While simulating the performance of the TFT and WSLS strategy, we
start by assuming that in the initial iteration all users are
defectors. At the end of the initial iteration we choose one user
at random to become a cooperator. 
The simulation results show that COOP yields a
minimal total energy consumption among all four strategies (which
is expected), followed by TFT, as shown in Table \ref{tab:mean_std}. 
\begin{figure}[!h]
\begin{center}
\includegraphics[scale=0.36]{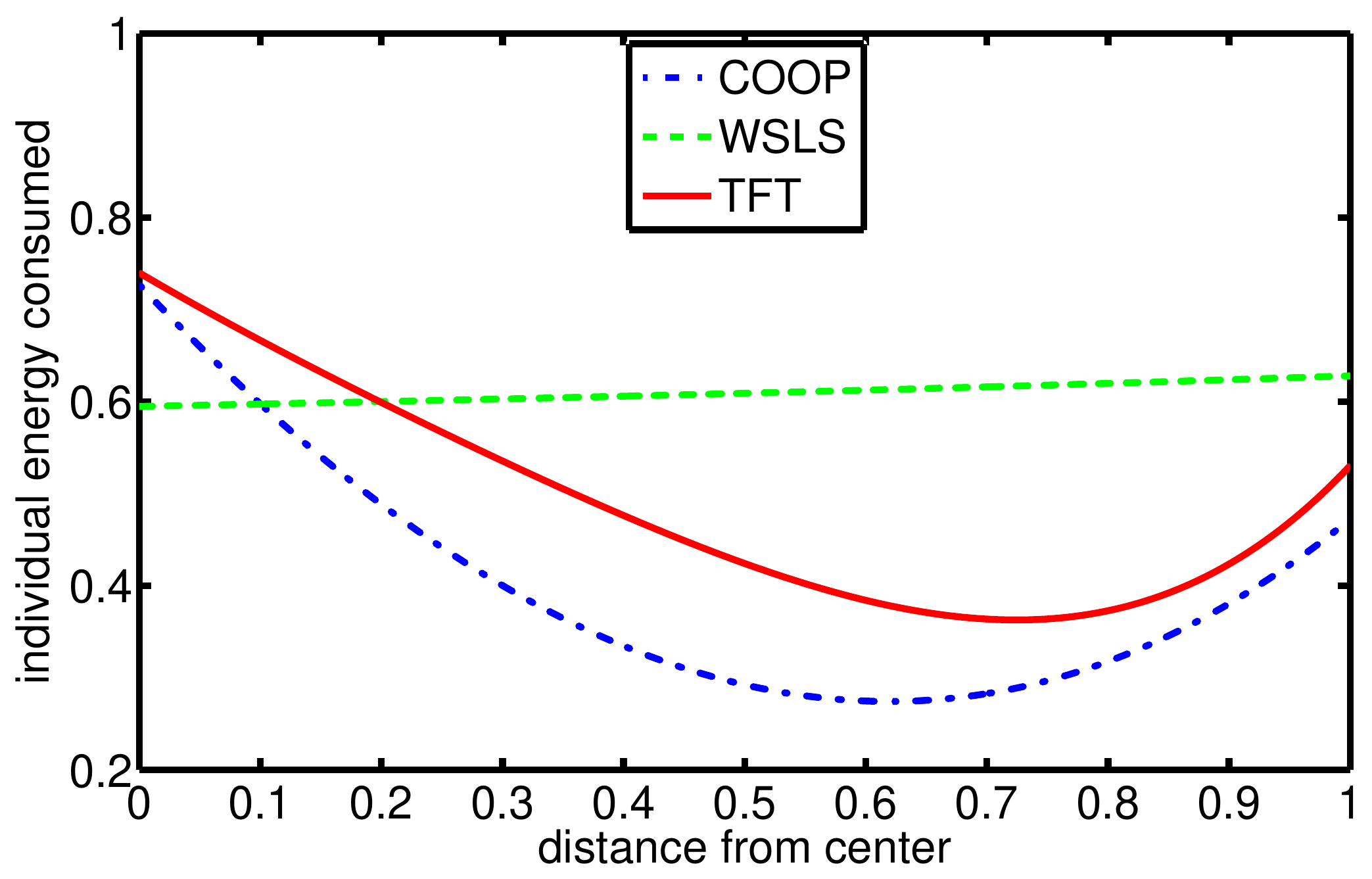}
\end{center}
\caption{Individual energy consumption with COOP, TFT and WSLS.}
\label{fig:energy_spent_vs_r_wout_def}       
\end{figure}

However, the results
also indicate that the COOP might not be the optimal strategy from
the perspective of the energy consumption of the individual nodes, as Fig.~\ref{fig:energy_spent_vs_r_def_coop} and Fig.~\ref{fig:energy_spent_vs_r_wout_def} indicate. For example,
when the network nodes follow the WSLS strategy, the nodes closer
to the center are characterized by lower individual energy
consumption compared to the other strategies. Further, the WSLS
strategy yields a more balanced energy consumption as function of
the geographical distribution of the nodes, as presented in
Table~\ref{tab:mean_std}.

\subsection{Simulations: Network with Central Infrastructure Node}
The second network architecture we
address corresponds to a cellular (infrastructure) scenario as it
assumes that the users located in one area transmit their signals
to a central infrastructure node, e.~g. access point, relay, or
base station.
\subsubsection{Cooperative behaviour}
Due to the specific network geometry (central infrastructure node
which receives all packets), the mechanism of spreading of
cooperation in this network shows significant difference with the
wireless ad hoc network. To see why this is the case, let us take
that a cooperator is placed at distance $r_0$ from the center. Due
to the specific network configuration, only nodes which are at
distance $r>r_0$ (i.~e. further away from the center) can benefit
from the cooperative act. As result, only these nodes can change
their behavior from defectors to cooperators. In other words,
spreading of cooperation is sensitive to the location of the
initial cooperator. This is a major difference from the previous
case, where no such constraint has been observed.

\begin{figure}[!htb]
\begin{center}
\includegraphics[scale=0.36]{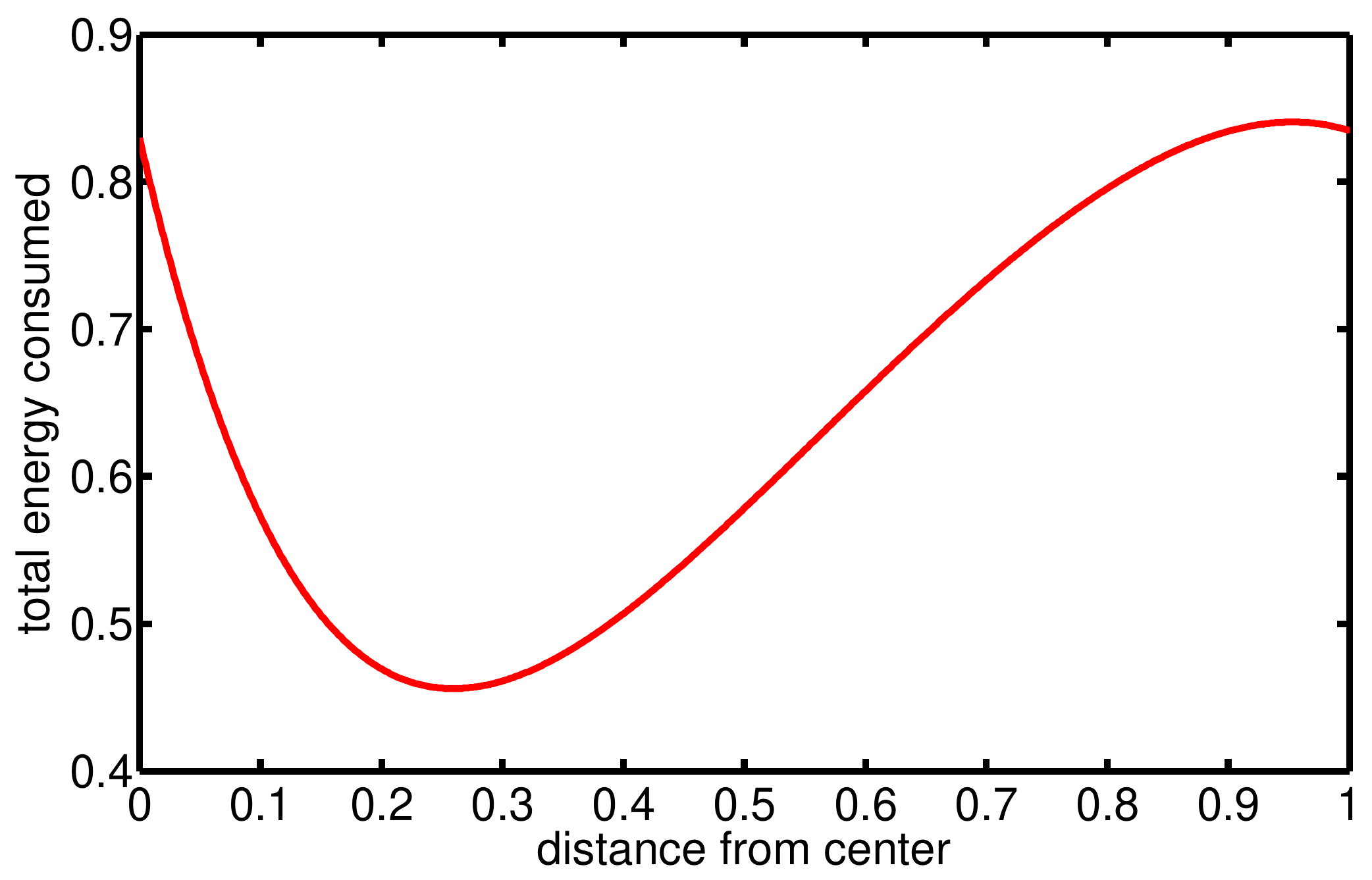}
\caption{Total energy consumed with TFT depending on the location of the initial cooperator}
\label{fig:energy_by_frist_coop}
\end{center}
\end{figure}

This observation has the implication that, in order to quantify
correctly the effect of cooperation on the energy consumption in
the network, one has to average over all different placements of
the initial cooperator. Fig.~\ref{fig:energy_by_frist_coop}
depicts the total energy consumption for the TFT strategy\footnote{Since the cooperative behaviour vanishes in the case of the WSLS strategy, only the result for the TFT strategy is illustrated.}(averaged over the number of nodes), 
presented as a function of
the location of the initial cooperator. We observe that the
overall energy consumption is minimal when the initial cooperator
is approximately at distance $r_0=0.26$ from the center. This behavior is somewhat surprising, since
one would expect that the most beneficial set up is when the initial
cooperator is located as close as possible to the center.

\subsubsection{Energy consumption}

As result of the network configuration, the behaviour of the TFT
and the WSLS strategy in this network architecture shows
significant difference. The most important observation is that
when the nodes adopt the WSLS strategy, cooperation does not
persist in the network on long term. Namely, after the
introduction of the initial cooperator, the cooperation behavior
spreads in the network, but fades out over a finite number of
iterations. In the case when the nodes adopt the TFT strategy, on
the other hand, simulation results show that cooperation persists
over time (although the region where cooperators are found)
depends on the placement of the initial cooperator. The TFT
strategy also yields a fairly low total energy consumption in the
network (second best, right after the COOP strategy), as shown in  Table~\ref{tab:mean_std_center}.

\begin{figure}[!h]
\begin{center}
\includegraphics[scale=0.36]{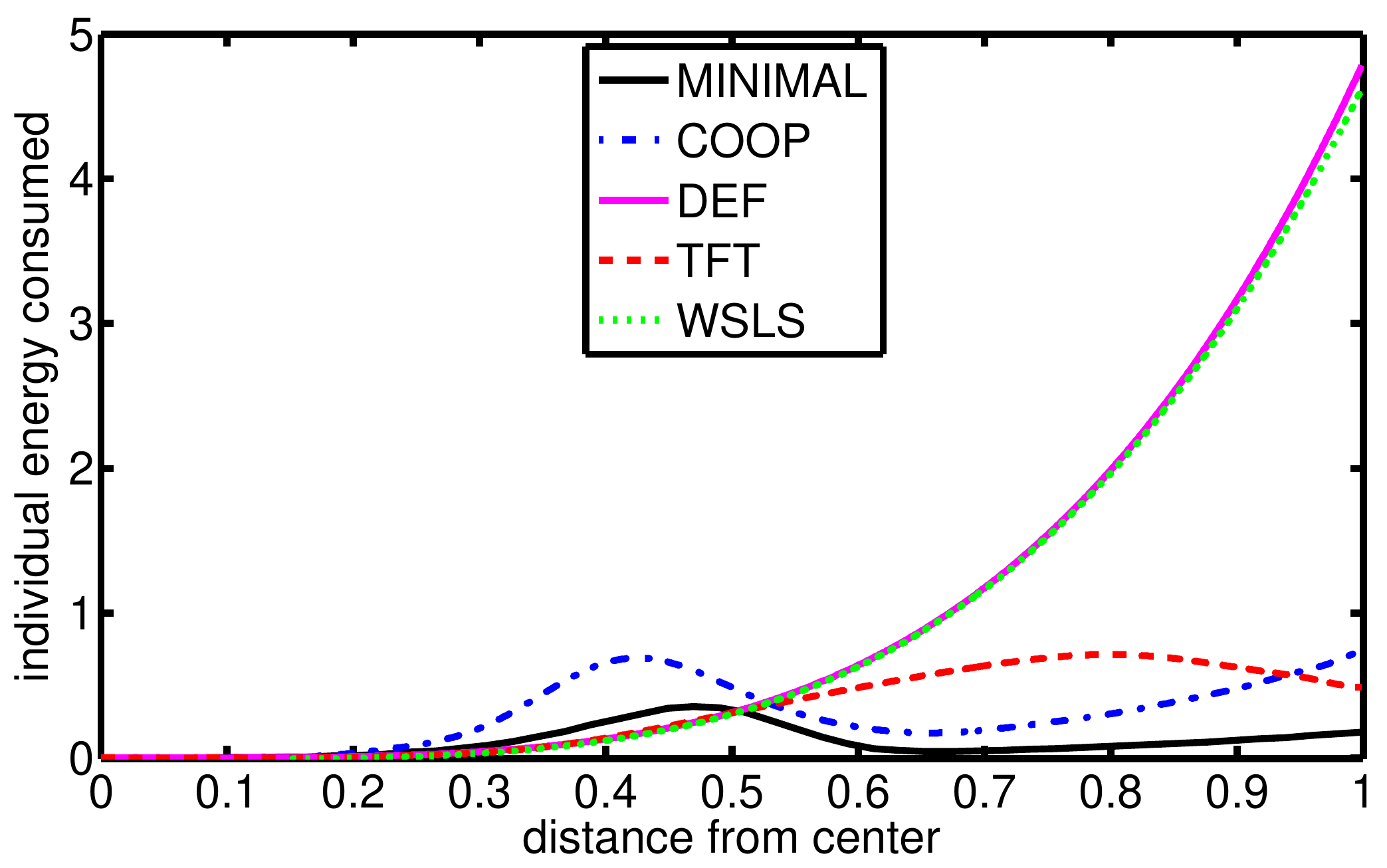}
\caption{Individual energy as function of the distance from the center}
\label{fig:individual_energy_all_center}
\end{center}
\end{figure}

\begin{savenotes}
\begin{table}[!h]
\caption{Mean and standard deviation of the energy $\mathcal{E}$ in the cellular scenario}
\label{tab:mean_std_center}
\begin{center}
\begin{tabular}{ l | r | r }
  strategy & mean($\mathcal{E}$) & std($\mathcal{E}$) \\
  \hline
   DEF      &  1.00000  &  1.34555  \\
   COOP     &  0.29348  &  0.78187  \\
   MINIMAL &  0.11021  &  0.92291  \\
   TFT\footnote{The total energy consumption depends on the location of the individual cooperator. The values in the table are shown for cooperator locations which lead to the minimal possible energy consumption.}      &  0.46329  &  0.81327  \\
   WSLS     &  0.99534  &  1.33419  \\
\end{tabular}
\end{center}
\end{table}
\end{savenotes} 
%
%
Additionally, it can be observed that WSLS and DEF have almost identical energy consumption.
This is expected, since after the initial cooperation frenzy, the number of defectors gradually increases leading to extinction of cooperation. The individual energy consumption as function of the distance from the center, when different strategies are applied, is displayed in Fig.~\ref{fig:individual_energy_all_center}. When the nodes act accordingly to the WSLS strategy, the energy consumption increases polynomially with the distance, due to the extinction of cooperation. The energy consumption with TFT increases gradually up to $0.8$ distance units from the center, and decreases for the nodes over this point. We can conclude that TFT leads to a reduction in the individual energy consumption for all nodes when compared to DEF. This is different from the COOP strategy where some nodes (for example those at distance $0.4$) spend more energy as cooperators as opposed to when they are defectors.

\subsection{A special case: dense homogeneous network with central infrastructure node and full cooperation}
This case addresses the scenario where the network is dense
enough so that at least one cooperator is present in every
(sufficiently small) area. This assumption is not only convenient
as it simplifies the analysis, but is also realistic to some
extent since it covers the case of large networks spreading over a region of fixed area. Additionally, we consider that all nodes are cooperators.
Hence, the performance of the system 
in this setting can be seen
as the ultimate performance bound for any strategy (or a set of
mixed strategies) in terms of the total energy consumption in the
network.  
\begin{figure}[!htb]
\begin{center}
\includegraphics[scale=0.29]{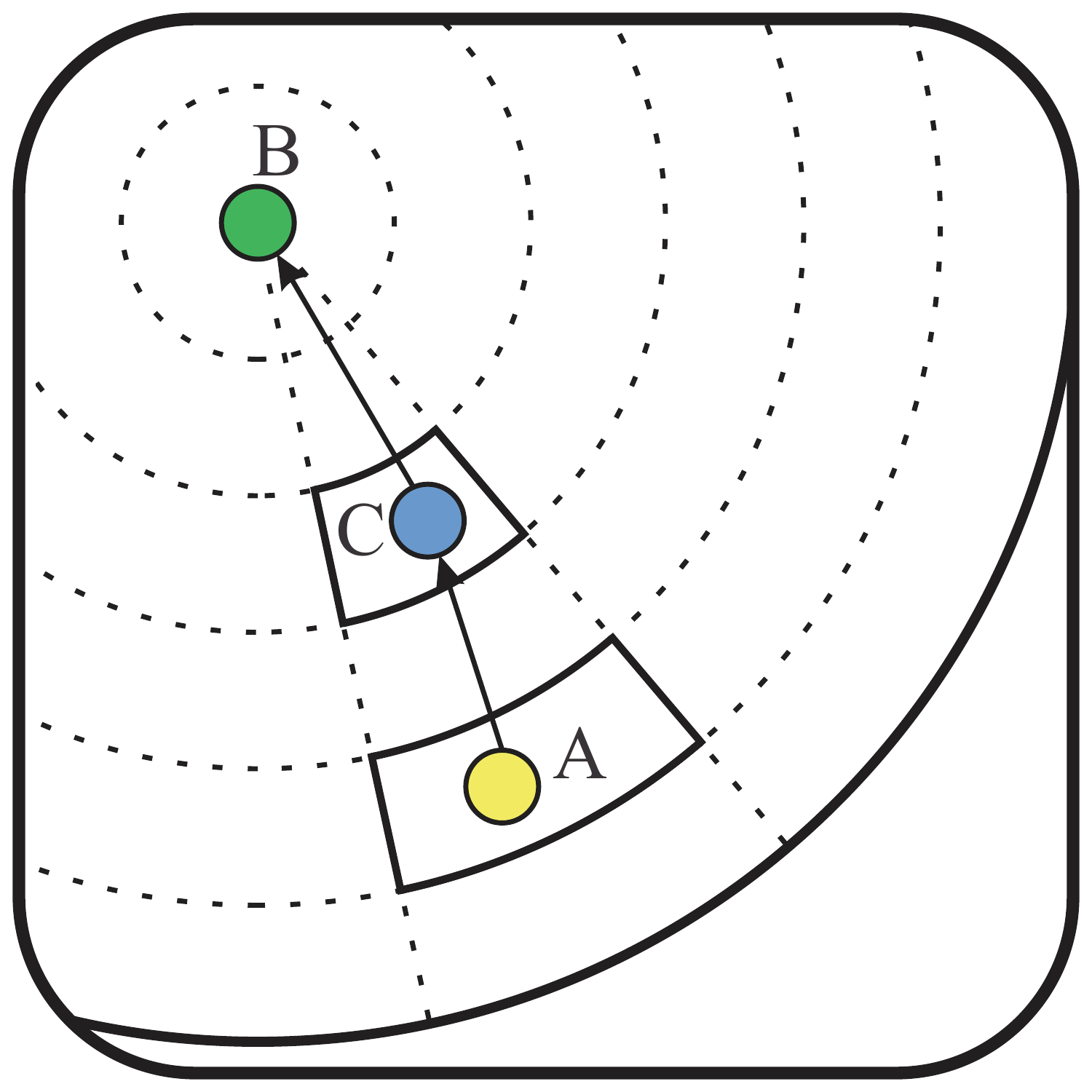}
\caption{Example of the communication in a cellular network}
\label{fig:prsteni_view}
\end{center}
\end{figure}

In a typical
communication scenario, node A (located at distance $x$ from the center) 
transmits to the central node B with the help of a 
cooperator C (located at distance $y<x$ from B). 
The high-density assumption allows to
assume that the node C lies on the line AB, or at least very
close to it (see Fig.~\ref{fig:prsteni_view}). 
In order to distribute the cost of cooperation, we assume that
A can choose among all cooperators between it and B (which
are close to the line AB), with some probability. Let the
probability density function for choosing (targeting) the cooperator at
distance $y$, when the sender is at distance $x$, be $P(x,y)$. The
average transmission energy for a node at distance $r$ from the
central node thus reads
\begin{equation}
\mathcal{E}_T(r) = K\int_0^r P(r,y)(r-y)^{\alpha} dy,
\label{eq:Transmission_Energy}
\end{equation}
where $K$ is a normalization constant which accounts for all
propagation factors, apart from the signal path loss.

The density $P(x,y)$ corresponds to all pairs
transmitter-cooperator which are located in some infinitesimal
regions (see Fig.~ \ref{fig:prsteni_view}), where the area of such
infinitesimal region is proportional to its distance from the
central node. We assume that the retransmissions of messages
originating from the infinitesimal area around A are shared
equally by the cooperators from the infinitesimal area around C
(we consider only the messages targeting the cooperators in this
area according to the probability distribution). From the proportionality of the areas, it follows that the average
number of messages (per node) to be retransmitted by a cooperator
C at distance $y$ from the central node, and originating from a
sender A at distance $x$, is larger for a factor $x/y$ than the
average number of generated messages (targeting at C\rq{}s
neighbourhood). Hence, the average energy spent for
cooperation of a node at distance $r$ from the central node reads
\begin{equation}
\mathcal{E}_C(r) = K\int_r^R P(x,r) r^{\alpha} \frac{x}{r} dx.
\label{eq:Cooperation_Energy}
\end{equation}
The average energy consumed by a node at distance $r$ from the
central node (including both the energy for transmission of own
messages and the energy for cooperation) is given by
\begin{equation}
\mathcal{E}(r) = K \int_0^r P(r,y)(r-y)^{\alpha} dy + K \int_r^R
P(x,r)r^{\alpha} \frac{x}{r} dx. \label{eq:Energy_Distance_r}
\end{equation}

By using (\ref{eq:Energy_Distance_r}), one can calculate the
total energy consumption in the network as
\begin{align}
\mathcal{E}_{\rm total} 
= K \int_0^R dx \int_0^x P(x,y) Q(x,y) dy, \label{eq:Total_Energy}
\end{align}
where
\begin{equation}
Q(x,y) = (x-y)^{\alpha} + y^{\alpha-1} x.
\end{equation}

Ideally, the optimal choice of the distribution $P(x,y)$ would 
minimize the total energy consumption and balance the
individual energy consumption simultaneously. However, since both
requirements might be contradictory, we first address each issue
separately.
\paragraph*{Minimizing the total energy consumption} Here we
derive the distribution of $P(x,y)$ which minimizes
$\mathcal{E}_{\rm total}$, and also present the resulting minimal
energy.
\begin{lemma}
\label{lemma:Minimal_Energy} The distribution $P(x,y)$ which
minimizes $\mathcal{E}_{\rm total}$ is given by
\begin{equation}
 P(x,y)  = \delta \left(y - y(q(x))\right),
\label{eq:Singular_P}
\end{equation}
where $Q_{\rm min}(x,y) \doteq q(x)$ denotes the minimal value of
$Q$ for fixed $x$ -- the minimization is done by varying $y$, and
$\delta(\cdot)$ is the Dirac $\delta$ function.
\end{lemma}
\begin{IEEEproof}
The proof is given in Appendix~\ref{sec:appendix_1}.
\end{IEEEproof}
Having the optimal $P(x,y)$, the minimal total energy
reads
\begin{align}
\mathcal{E}_{\rm min}&= K \int_0^R dx \int_0^x \delta \left(y - y(q(x))\right) Q(x,y) dy \nonumber \\
&= K \int_0^R q(x) dx. \label{eq:Minimal_Energy}
\end{align}
Clearly, the last integral depends on $\alpha$, since $q(x)$ depends on $\alpha$, $q(x) = q_{\alpha}(x)$.
Singularity of distributions implies that the transmitters from
the ring with radius $x$ should send their messages to their peers
at the ring with radius $y(q(x))$. We denote the cooperation strategy associated 
strategy with the minimal total energy consumption as MINIMAL. The total energy consumption of this strategy is illustrated in Table~\ref{tab:mean_std_center}.
\paragraph*{Balancing the individual energy consumption}
While the distribution (\ref{eq:Minimal_Energy}) minimizes the total energy consumption in the network, 
it is not the optimal solution when it comes to the energy consumption 
of the individual nodes. Indeed, as Fig.~\ref{fig:individual_energy_all_center} shows, the individual energy consumption depends on the node location. For example, nodes at distance $0.5$ (approximately) from the center consume more energy than other nodes. Perfect balancing would require that the distribution
$P(x,y)$ is such that the individual energy consumption is
independent on the distance $r$
\begin{align}
\mathcal{E}(r)&= K \int_0^r P(r,y)(r-y)^{\alpha} dy + K \int_r^R
P(x,r) r^{\alpha} \frac{x}{r} dx\nonumber\\
&=const. \label{eq:Constant_Energy}
\end{align}
Analytical solution of (\ref{eq:Constant_Energy}) (if it exists at all),
seems to be out of reach. In order to simplify the analysis, one
can relax the demand for equal energy spending and search for a
solution with \textit{as balanced as possible} individual consumption. This means
finding the coordination pattern among the nodes which leads to
smallest possible variation of the consumed energy. To make the
problem more tractable one can divide the circle into $N$ rings of width $r$, such that $R=Nr$. This discretization of the
problem would allow for deployment of some optimization
techniques. The directions for solving the problem in the discrete
case are presented in Appendix \ref{sec:Appendix_2}.

\section{Discussion}
\label{sec:discussion}
\subsection{Interpretation of the results}
The results clearly indicate that cooperation
can spread in wireless networks, even in networks with selfish
nodes which adopt simple strategies and update their behavior only
based on the individual fitness. In addition, there are several
other important conclusions, which arise as result of the analysis.
\paragraph*{Spreading and persistence of cooperation is
affected by network architecture} In the ad-hoc network setup the
adoption of the WSLS strategy results in spreading of the
cooperative behavior in the network. Cooperation remains stable
over time, resulting in decreased energy consumption. In the
network with a central infrastructure node, on the other hand, the
adoption of the WSLS strategy does not favor cooperation on long
run-after the initial spread, cooperation fades out with time. The
cooperative behavior in the network where the nodes adopt the TFT
strategy, on the other hand, seems to be more robust to different
network topologies, as it persists over time in both network
setups. There is additional difference between the network
architectures, since the spreading of cooperation shows to be
sensitive to the location of the initial cooperator in the network
with central infrastructure node, in contrast to the ad-hoc
network setup.
\paragraph*{Cooperation has positive effect on the total
energy consumption in the network} When cooperation persists, both
TFT and WSLS yield significant decrease in the energy consumption
compared to the case without cooperation (DEF strategy). The
performance is, as expected upper-bounded by the performance of
the COOP strategy. However, the COOP strategy requires mechanisms
to enforce cooperation in order to ensure that all nodes
cooperate, which are particularly difficult to implement in
decentralized, heterogeneous networks.
\paragraph*{Different strategies balance the individual
energy spending in a different way} While cooperation has positive
effect on the average energy consumption, the individual energy
consumption depends on the location of the individual nodes.
Moreover, different strategies balance the total energy spending
over the individual nodes in a different way. For example, in the
case of the ad-hoc network, the WSLS strategy yields lower energy
consumption then TFT and COOP for the nodes near the center.
Additionally, while WSLS yields higher total energy consumption
compared to TFT and COOP, the consumption of the individual nodes
is approximately constant over different locations (distances from
the center). This behavior of WSLS could be particularly relevant
in the setup where the network nodes have finite energy capacity
(finite buffer). In the network with a central infrastructure
node, the COOP strategy seems to increase the cooperative burden
for the nodes located in the middle between the central node and
the peripheral nodes, while the TFT strategy seems to share the
cost of cooperation more equally among the cooperative nodes, as
indicated by the simulation results.
\subsection{Relevance to systems other than communication networks}
Although this work addresses energy efficiency in wireless communication systems, 
it is possible that the results have implications on the understanding of certain biological and social 
phenomena. We recall that the main feature of wireless networks which justifies cooperative behaviour is the fact that the cost of communication (transmission) is a function which is polynomial in the distance. However, similar functions (although not always well understood) may be objects to optimization in biological systems as well. As an example we can take the neural system, where neurons differentiate from multipotent stem cells and migrate to their final residence in the system. When these neurons reach their residence, they extend an axon which needs to travel a certain distance to attach to other neurons (forming a synapse) and enable inter-neuronal communication. Since axons may travel long distances, axonal trajectories appear to be broken up into a series of smaller movements, where the axon finds intermediate targets that act as choice points (also called guidepost cells) \cite{shen2004synaptic}. This behaviour has clear analogy in the relaying (retransmissions) addressed in the wireless network scenario. In this particular example, the insights gained from the analysis of wireless networks where the behaviour of the nodes is determined by evolutionary-like rules, could be applied to access the benefits and costs associated with growing axons directly to the target, and using intermediate targets, i.~e. guidepost cells.

\section{Conclusions and Future Work}
\label{sec:conclusion} We investigated the mechanisms for
promotion of cooperation in decentralized wireless networks. The
approach was motivated by recent results in evolutionary biology
which suggest that cooperation can be favoured by natural
selection, if a certain mechanism is at work. We modelled the
wireless network as a graph, where benefits and costs were
associated with the strategy that the network users follow. In
game-theoretic spirit, the nodes based their behavior on
calculations of their energy spending. We presented numerical study of cooperative communication
scenarios based on simple local rules, which is in contrast to
most of the approaches in the literature which enforce cooperation
by using complex algorithms and require strategic complexity of
the network nodes. 
The simulations show that even selfish decision making (such as one based on TFT or WSLS strategy) 
of the nodes can lead to emergence of cooperation. These
observations serve as indicator that uncomplicated local rules,
followed by simple fitness evaluation, can generate network
behavior which yields global energy efficiency. We identify several major directions for future work, as formulated in the following.
\paragraph*{Individual strategy selection} We recall that in this work we adopted the convention that the same
strategy was used by all users in all iterations. In a future
version of the work, we will consider the case where each of the
individual users is allowed to choose its own strategy at every
iteration. As discussed, the results from the simulations indicate
that, depending on the node distance from the center, distinct
nodes could find optimal to follow different strategies. It is
expected that this analysis will bring valuable insights in the
dependencies between the choice of optimal strategy for the
individual users and the network topology.
\paragraph*{Nodes with finite energy buffers, energy harvesting}
In addition, it will be interesting to evaluate the network
behaviour in the case when the nodes have buffers with limited
energy capacity, under a particular random arrival process. This
is in contrast to the here addressed scenario which assumes
nodes with infinite-length buffers. We expect that the
adoption of this more realistic assumption will influence both the
behaviour of the individual nodes and the way energy is consumed in
the network. Additionally, it is expected that will yield qualitatively different performance of 
the here addresses strategies, as indicated in Section \ref{sec:discussion}. This more general approach also includes the energy
harvesting scenario where the nodes harvest energy quanta from the
environment according to some arrival process.
\paragraph*{Implication on other systems}
While this work was motivated by observations in biological systems and social systems, 
it is possible that the results have implications on the understanding of certain biological and social 
phenomena. This would establish a connection in the other direction, where lessons from artificial systems (such as wireless communication networks) can be applied to natural systems. 
        
\appendices

\section{Proof of Lemma~\ref{lemma:Minimal_Energy}}
\label{sec:appendix_1}
Since the function $Q(x,y)$ is non-negative, the following
inequality holds
\begin{equation}
\int_0^R dx \int_0^x P(x,y) Q(x,y) dy \geq \int_0^R dx \int_0^x
P(x,y) q(x) dy, \label{eq:Inequality}
\end{equation}
where $q(x)\doteq Q_{\rm min}(x,y)$ denotes the minimal value of
$Q$ for fixed $x$ -- the minimization is done by varying $y$. The
right hand side of the last inequality then simplifies
\begin{align}
\int_0^R dx \int_0^x P(x,y) q(x) dy = \int_0^R q(x) dx,
\label{eq:Inequality_Integral}
\end{align}
due to the normalization of the distribution, $\int\int P(x,y)dx
dy=1$. The equality in (\ref{eq:Inequality}) will hold only if one
chooses a singular distribution
\begin{equation}
 P(x,y)  = \delta \left(y - y(q(x))\right),
\label{eq:Singular_P1}
\end{equation}
located at the point $y(q(x))$ which corresponds to the minimum of
the function $Q(x,y)$ for fixed $x$.

\section{Balancing the individual energy consumption: Discrete
approximation}
\label{sec:Appendix_2}
When the number of rings $N$ is large, for the purpose of energy calculation, 
one can assume that the nodes are located in the
middle of the rings. Furthermore, we assume that every node sends its
message with some probability to some peer in some inner ring, or
with some probability directly to the relay. For the node in the
ring $i$ the set of those probabilities is $p_{i,j}$, where
$j=0,1,...i-1$, and $p_{i,0}$ is the probability for sending the
message to the relay. The energy for transmission for that node
will be
\begin{equation}
\mathcal{E}_t(i) = \sum_{j=0}^{i-1} p_{i,j} [r(i-j)]^{\alpha} =
r^{\alpha} \sum_{j=0}^{i-1} p_{i,j} (i-j)^{\alpha}.
\label{eq:Transmission_Discrete}
\end{equation}
On the other hand, the cooperation will consume energy
\begin{equation}
\mathcal{E}_c(i) = \sum_{k=i+1}^{N} p_{k,i} (ir)^{\alpha}
\frac{k}{i} = r^{\alpha} \sum_{k=i+1}^{N} p_{k,i}
i^{\alpha}\frac{k}{i}, \label{eq:Cooperation_Discrete}
\end{equation}
where the factor $k/i$ appears for the same reason as in
(\ref{eq:Cooperation_Energy}). Total energy thus consists of two
sums
\begin{equation}
\mathcal{E}(i) = r^{\alpha} \left[\sum_{j=0}^{i-1} p_{i,j}
(i-j)^{\alpha} + \sum_{k=i+1}^{N} p_{k,i} i^{\alpha}\frac{k}{i}
\right]. \label{eq:Total_Discrete}
\end{equation}
The last expression could be written more neatly as
\begin{equation}
\mathcal{E}(i) = r^{\alpha} \sum_{j=0}^{N} (\beta_{i,j} p_{i,j} +
\gamma_{j,i} p_{j,i} ), \label{eq:Total_Discrete_Simplified}
\end{equation}
where the non-zero coefficients are $\beta_{i,j}=(i-j)^{\alpha}$
for $j=0,1,...,i-1$ and $\gamma_{j,i} = ji^{\alpha-1}$ for $j=i+1,
i+2, ...,N$. This means that for particular $i$ and $j$ only one
of  $\beta_{i,j}$ and $\gamma_{j,i}$ is non-zero. The average
energy is
\begin{eqnarray}
\langle \mathcal{E}\rangle = \frac{r^{\alpha}}{N} \sum_{i=1}^{N} \mathcal{E}(i) &=& \frac{r^{\alpha}}{N}  \sum_{i=1}^{N} \sum_{j=0}^{N} (\beta_{i,j} p_{i,j} + \gamma_{j,i} p_{j,i} ) \nonumber \\
&=&\frac{r^{\alpha}}{N}  \sum_{i=1}^{N} \sum_{j=0}^{i-1}
\delta_{i,j} p_{i,j}, \label{eq:Total_Discrete_Simplified_1}
\end{eqnarray}
where $\delta_{i,j} = \beta_{i,j} + \gamma_{j,i} $. As a measure
of the balance is the mean squared deviation
\begin{equation}
\sigma^2 = \langle (\mathcal{E}- \langle \mathcal{E}\rangle )^2
\rangle = \frac{1}{N} \sum_{j=1}^{N} [\mathcal{E}(i)- \langle
\mathcal{E}\rangle]^2. \label{eq:Variance}
\end{equation}
Since the optimal parameters $p_{i,j}$ are probabilities, the
normalization puts a constraint
\begin{equation}
\sum_{j=0}^{N} p_{i,j} = \sum_{j=0}^{i-1} p_{i,j} = 1.
\label{eq:constraint}
\end{equation}
Hence, in the discrete case, if one searches for a solution which balances the individual energy consumption in the network, one should minimize (\ref{eq:Variance}), subject to the constraint (\ref{eq:constraint}).This can be performed, for example, by using certain modeling systems for convex optimization. 

\bibliographystyle{IEEEtran}
\bibliography{references}

\begin{thebibliography}{10}
\providecommand{\url}[1]{#1}
\csname url@samestyle\endcsname
\providecommand{\newblock}{\relax}
\providecommand{\bibinfo}[2]{#2}
\providecommand{\BIBentrySTDinterwordspacing}{\spaceskip=0pt\relax}
\providecommand{\BIBentryALTinterwordstretchfactor}{4}
\providecommand{\BIBentryALTinterwordspacing}{\spaceskip=\fontdimen2\font plus
\BIBentryALTinterwordstretchfactor\fontdimen3\font minus
  \fontdimen4\font\relax}
\providecommand{\BIBforeignlanguage}[2]{{%
\expandafter\ifx\csname l@#1\endcsname\relax
\typeout{** WARNING: IEEEtran.bst: No hyphenation pattern has been}%
\typeout{** loaded for the language `#1'. Using the pattern for}%
\typeout{** the default language instead.}%
\else
\language=\csname l@#1\endcsname
\fi
#2}}
\providecommand{\BIBdecl}{\relax}
\BIBdecl

\bibitem{bandyopadhyay2005spatio}
S.~Bandyopadhyay, Q.~Tian, and E.~J. Coyle, ``Spatio-temporal sampling rates
  and energy efficiency in wireless sensor networks,'' \emph{IEEE/ACM
  Transactions on Networking (TON)}, vol.~13, no.~6, pp. 1339--1352, 2005.

\bibitem{sendonaris2003user1}
A.~Sendonaris, E.~Erkip, and B.~Aazhang, ``User cooperation diversity, part i:
  System description,'' \emph{IEEE Transactions on Communications}, vol.~51,
  no.~11, pp. 1927--1938, 2003.

\bibitem{sendonaris2003user2}
------, ``User cooperation diversity, part ii: Implementation aspects and
  performance analysis,'' \emph{IEEE Transactions on Communications}, vol.~51,
  no.~11, pp. 1939--1948, 2003.

\bibitem{cadambe2008}
V.~R. Cadambe and S.~A. Jafar, ``Interference alignment and degrees of freedom
  {of the $K$-user} interference channel,'' \emph{IEEE Transactions on
  Information Theory}, vol.~54, no.~8, pp. 3425--3441, 2008.

\bibitem{zhao2005energy}
Q.~Zhao and L.~Tong, ``Energy efficiency of large-scale wireless networks:
  proactive versus reactive networking,'' \emph{Selected Areas in
  Communications, IEEE Journal on}, vol.~23, no.~5, pp. 1100--1112, 2005.

\bibitem{el2006optimal}
A.~El~Gamal and J.~P. Mammen, ``Optimal hopping in ad hoc wireless networks.''
  in \emph{INFOCOM}, 2006.

\bibitem{zhong2003sprite}
S.~Zhong, J.~Chen, and Y.~R. Yang, ``Sprite: A simple, cheat-proof,
  credit-based system for mobile ad-hoc networks,'' in \emph{Twenty-Second
  Annual Joint Conference of the IEEE Computer and Communications, INFOCOM
  2003}, vol.~3.\hskip 1em plus 0.5em minus 0.4em\relax IEEE, 2003, pp.
  1987--1997.

\bibitem{buchegger2003wiopt}
S.~Buchegger, J.~Le~Boudec \emph{et~al.}, ``The effect of rumor spreading in
  reputation systems for mobile ad-hoc networks,'' in \emph{WiOpt'03: Modeling
  and Optimization in Mobile, Ad Hoc and Wireless Networks}, 2003.

\bibitem{liu2003reputation}
Y.~Liu and Y.~R. Yang, ``Reputation propagation and agreement in mobile ad-hoc
  networks,'' in \emph{Wireless Communications and Networking, WCNC 2003},
  vol.~3.\hskip 1em plus 0.5em minus 0.4em\relax IEEE, 2003, pp. 1510--1515.

\bibitem{anantvalee2007reputation}
T.~Anantvalee and J.~Wu, ``Reputation-based system for encouraging the
  cooperation of nodes in mobile ad hoc networks,'' in \emph{IEEE International
  Conference on Communications, ICC'07}.\hskip 1em plus 0.5em minus 0.4em\relax
  IEEE, 2007, pp. 3383--3388.

\bibitem{mundinger2008analysis}
J.~Mundinger and J.-Y. Le~Boudec, ``Analysis of a reputation system for mobile
  ad-hoc networks with liars,'' \emph{Performance Evaluation}, vol.~65, no.~3,
  pp. 212--226, 2008.

\bibitem{marbach2005cooperation}
P.~Marbach and Y.~Qiu, ``Cooperation in wireless ad hoc networks: A
  market-based approach,'' \emph{IEEE/ACM Transactions on Networking (ToN)},
  vol.~13, no.~6, pp. 1325--1338, 2005.

\bibitem{buttyan2001nuglets}
L.~Buttyan and J.-P. Hubaux, ``Nuglets: a virtual currency to stimulate
  cooperation in self-organized mobile ad hoc networks,'' 2001.

\bibitem{srinivasan2005analytical}
V.~Srinivasan, P.~Nuggehalli, C.~Chiasserini, and R.~R. Rao, ``An analytical
  approach to the study of cooperation in wireless ad hoc networks,''
  \emph{Wireless Communications, IEEE Transactions on}, vol.~4, no.~2, pp.
  722--733, 2005.

\bibitem{felegyhazi2006nash}
M.~Felegyhazi, J.-P. Hubaux, and L.~Buttyan, ``Nash equilibria of packet
  forwarding strategies in wireless ad hoc networks,'' \emph{Mobile Computing,
  IEEE Transactions on}, vol.~5, no.~5, pp. 463--476, 2006.

\bibitem{lai2008cooperation}
L.~Lai and H.~El~Gamal, ``On cooperation in energy efficient wireless networks:
  the role of altruistic nodes,'' \emph{Wireless Communications, IEEE
  Transactions on}, vol.~7, no.~5, pp. 1868--1878, 2008.

\bibitem{axelrod1981evolution}
R.~Axelrod and W.~D. Hamilton, ``The evolution of cooperation,''
  \emph{Science}, vol. 211, no. 4489, pp. 1390--1396, 1981.

\bibitem{nowak2006}
H.~Ohtsuki, C.~Hauert, E.~Lieberman, and M.~A. Nowak, ``A simple rule for the
  evolution of cooperation on graphs and social networks,'' \emph{Nature}, vol.
  441, no. 7092, pp. 502--505, 2006.

\bibitem{nowak2006five}
M.~A. Nowak, ``Five rules for the evolution of cooperation,'' \emph{Science},
  vol. 314, no. 5805, pp. 1560--1563, 2006.

\bibitem{cremer2012}
J.~Cremer, A.~Melbinger, and E.~Frey, ``Growth dynamics and the evolution of
  cooperation in microbial populations,'' \emph{Scientific reports 2, Article
  number 281}, no. doi:10.1038/srep00281, 2012.

\bibitem{lozano2012}
A.~Lozano, R.~W. Heath~Jr., and J.~G. Andrews, ``Fundamental limits of
  cooperation,'' \emph{http://arxiv.org/abs/1204.0011v1}, 2012.

\bibitem{soares2008cleaning}
M.~C. Soares, I.~M. C{\^o}t{\'e}, S.~Cardoso, and R.~Bshary, ``The cleaning
  goby mutualism: a system without punishment, partner switching or tactile
  stimulation,'' \emph{Journal of Zoology}, vol. 276, no.~3, pp. 306--312,
  2008.

\bibitem{mehdiabadi2006social}
N.~J. Mehdiabadi, C.~N. Jack, T.~T. Farnham, T.~G. Platt, S.~E. Kalla,
  G.~Shaulsky, D.~C. Queller, and J.~E. Strassmann, ``Social evolution: kin
  preference in a social microbe,'' \emph{Nature}, vol. 442, no. 7105, pp.
  881--882, 2006.

\bibitem{faaborg1995confirmation}
J.~Faaborg, P.~Parker, L.~DeLay, T.~De~Vries, J.~Bednarz, S.~M. Paz,
  J.~Naranjo, and T.~Waite, ``Confirmation of cooperative polyandry in the
  galapagos hawk (buteo galapagoensis),'' \emph{Behavioral Ecology and
  Sociobiology}, vol.~36, no.~2, pp. 83--90, 1995.

\bibitem{buchegger2002performance}
S.~Buchegger and J.-Y. Le~Boudec, ``Performance analysis of the confidant
  protocol,'' in \emph{Proceedings of the 3rd ACM international symposium on
  Mobile ad hoc networking \& computing}.\hskip 1em plus 0.5em minus
  0.4em\relax ACM, 2002, pp. 226--236.

\bibitem{michiardi2002core}
P.~Michiardi and R.~Molva, ``Core: a collaborative reputation mechanism to
  enforce node cooperation in mobile ad hoc networks,'' in \emph{Advanced
  Communications and Multimedia Security}.\hskip 1em plus 0.5em minus
  0.4em\relax Springer, 2002, pp. 107--121.

\bibitem{caire2004suboptimality}
G.~Caire, D.~Tuninetti, and S.~Verd{\'u}, ``Suboptimality of tdma in the
  low-power regime,'' \emph{Information Theory, IEEE Transactions on}, vol.~50,
  no.~4, pp. 608--620, 2004.

\bibitem{ozgur2011operating}
A.~{\"O}zg{\"u}r, O.~L{\'e}v{\^e}que, D.~Tse \emph{et~al.}, ``Operating regimes
  of large wireless networks,'' \emph{Foundations and Trends{\textregistered}
  in Networking}, vol.~5, no.~1, pp. 1--107, 2011.

\bibitem{lasaulce2011game}
S.~Lasaulce and H.~Tembine, \emph{Game theory and learning for wireless
  networks: fundamentals and applications}.\hskip 1em plus 0.5em minus
  0.4em\relax Academic Press, 2011.

\bibitem{zhao2009scalability}
S.~Zhao and D.~Raychaudhuri, ``Scalability and performance evaluation of
  hierarchical hybrid wireless networks,'' \emph{Networking, IEEE/ACM
  Transactions on}, vol.~17, no.~5, pp. 1536--1549, 2009.

\bibitem{shen2004synaptic}
K.~Shen, R.~D. Fetter, and C.~I. Bargmann, ``Synaptic specificity is generated
  by the synaptic guidepost protein syg-2 and its receptor, syg-1,''
  \emph{Cell}, vol. 116, no.~6, pp. 869--881, 2004.

\end{thebibliography}

\end{document}